\documentclass[12pt]{article}
\usepackage{comment}
\usepackage{amsmath}
\usepackage{amssymb}
\usepackage{graphicx}
\usepackage{here}
\usepackage{subcaption}
\usepackage{url}
\usepackage[sort&compress, numbers, merge]{natbib}
\usepackage{physics}

\usepackage{todonotes}

\usepackage[colorlinks=true, linkcolor=blue, citecolor=blue,
urlcolor=black]{hyperref}

\begin{document}

\def\ps{\mathbf{p}}
\def\PS{\mathbf{P}}

\baselineskip 0.6cm

\def\simgt{\mathrel{\lower2.5pt\vbox{\lineskip=0pt\baselineskip=0pt
           \hbox{$>$}\hbox{$\sim$}}}}
\def\simlt{\mathrel{\lower2.5pt\vbox{\lineskip=0pt\baselineskip=0pt
           \hbox{$<$}\hbox{$\sim$}}}}
\def\simprop{\mathrel{\lower3.0pt\vbox{\lineskip=1.0pt\baselineskip=0pt
             \hbox{$\propto$}\hbox{$\sim$}}}}
\def\tr{\mathop{\rm tr}}
\def\SU{\mathop{\rm SU}}

\begin{titlepage}

\begin{flushright}
IPMU20-0084
\end{flushright}

\vskip 1.1cm

\begin{center}

{\Large \bf 
Cosmological Constraint on Vector Mediator of Neutrino-Electron Interaction in light of XENON1T Excess
}

\vskip 1.2cm
Masahiro Ibe$^{a,b}$, 
Shin Kobayashi$^{a}$, 
Yuhei Nakayama$^{a}$ and
Satoshi Shirai$^{b}$
\vskip 0.5cm

{\it

$^a$ {ICRR, The University of Tokyo, Kashiwa, Chiba 277-8582, Japan}

$^b$ {Kavli Institute for the Physics and Mathematics of the Universe
 (WPI), \\The University of Tokyo Institutes for Advanced Study, \\ The
 University of Tokyo, Kashiwa 277-8583, Japan}
}

\vskip 1.0cm

\abstract{
Recently, the XENON1T collaboration reported an 
excess in the electron recoil energy spectrum.
One of the simplest new physics interpretation is 
a new neutrino-electron interaction mediated by a light vector particle.
However, for the parameter region favored by this excess, the constraints from the stellar cooling are severe.
Still, there are astrophysical uncertainties on those constraints.
In this paper, we discuss the constraint on the light mediator from 
the effective number of neutrino 
$N_\mathrm{eff}$ in the CMB era, which provides an independent constraint.
We show that  $N_{\mathrm{eff}}$ is significantly enhanced and exceeds the current constraint
in the parameter region
favored for the XENON1T excess.
As a result, the interpretation by a light mediator heavier than about $1$\,eV is excluded by the $N_{\mathrm{eff}}$ constraint.
}

\end{center}
\end{titlepage}

\section{Introduction}
The XENON1T experiment has recently reported an excess in the electron recoil energy spectrum above the known expected background spectrum~\cite{Aprile:2020tmw}. Although the XENON1T experiment has not excluded unknown backgrounds, such as the $\beta$-decay of the tritium, the report is intriguing and has prompted many new physics interpretations. 

Among various new physics, one of the simplest possibilities is to introduce non-standard neutrino-electron interactions so that the solar neutrino flux explains the excess~\cite{DiLuzio:2020jjp,Boehm:2020ltd,Amaral:2020tga,Bally:2020yid,AristizabalSierra:2020edu,Khan:2020vaf,Lindner:2020kko}. For example, the XENON1T has discussed the sizable neutrino magnetic moment as a candidate for such a non-standard interaction. However, the best fit value is in tension with constraints from the white dwarfs and the globular clusters~\cite{daz2019constraint}.  Refs.\,\cite{Boehm:2020ltd,Amaral:2020tga,Bally:2020yid,AristizabalSierra:2020edu,Khan:2020vaf,Lindner:2020kko} have discussed new neutrino-electron interactions mediated by a light mediator,
where the mediator with a mass 
below $\order{100}$\,keV and 
the neutrino-electron coupling, $(g_eg_\nu)^{1/2}\sim 3\times 10^{-7}$ explains the excess.
The effects on the stellar cooling severely constrain the coupling constant of the light mediator~\cite{Grifols:1986fc,Grifols:1988fv,An_2013,Redondo_2013,Chang:2016ntp,Chang:2018rso}. 
However, those astrophysical constraints are still under debate as there are several uncertainties in the case of the mediator heavier than $1$\,eV~\cite{Redondo:2008aa,An_2013}.
For example, 
a new mediator produced inside the astronomical objects is reabsorbed before exiting the objects
 in the case of a large coupling to electron, which weakens the constraints. 

In this paper, we discuss a new constraint on the light mediator from the effective number of neutrino degrees of freedom, $N_{\mathrm{eff}}$, measured by the cosmic microwave background (CMB) observations.
In the standard cosmology, it is predicted that $N_{\mathrm{eff}}^{(\mathrm{SM})} \simeq 3.045$~\cite{deSalas:2016ztq,Akita:2020szl},
which is consistent with the current CMB measurement.
The introduction of 
new physics alters
the $N_\mathrm{eff}$
prediction, which provides a constraint
independent from the astrophysical constraints.
As the light mediator interpretation of the XENON1T excess requires a particle lighter than $\order{100}$ keV, the mediator mainly decays into the neutrinos. As we will see, the presence of the mediator significantly enhances $N_{\mathrm{eff}}$ even for
a very tiny coupling $g\sim 10^{-10}$.
The parameter region favored for the XENON1T excess results in $N_{\mathrm{eff}} > 5$ for $m_{Z'} \ge 1$\,eV, which exceeds
the upper limit of the Planck CMB only (joint Planck+BAO) 
constraint, $N_{\rm eff} = 2.92^{+0.36}_{-0.37}$  ($ 2.99^{+0.34}_{-0.33}$) at 95\% C.L.~\cite{Aghanim:2018eyx}.

The organization of the paper is as follows. In Sec.\,\ref{sec:model}, we explain the setup of the phenomenological model for the light mediator interpretation of the XENON1T excess. In Sec.\,\ref{sec:Boltzmann}, we show the full Boltzmann equation of the momentum distribution of the light mediator. In Sec.\,\ref{sec:Neffconstraint}, we obtain the constraints on the neutrino and the electron coupling of the light mediator. The final section is devoted to our conclusions. 

\section{Setup}
\label{sec:model}
As a phenomenological setup, we consider a light vector boson which couples to the neutrinos $\nu_i$ and the charged leptons $\psi_{i}$,
\begin{align}
\label{eq:model}
\mathcal{L}_{Z'} = g_{\nu}^{ij} Z'_{\mu}\nu^\dagger_{L,i} \Bar{\sigma}^\mu \nu_{L,j} + g_\ell Z'_\mu \bar{\psi}_{\ell}\gamma^\mu \psi_{\ell}\ ,
\end{align}
where $i,j,\ell = e, \mu, \tau$, $\nu_{L,i}$ is a left-handed 2-component Weyl spinor and $\psi_\ell$ is a 4-component Dirac spinor.%
\footnote{Depending on the ultraviolet completion of the massive gauge boson, there can be higher dimensional operators.
In the following analysis, we neglect such interactions.
}
The new vector mediator $Z'$ has a mass $m_{Z'}$.
Hereafter, we assume that the mediator is lighter than an electron-positron pair, $m_{Z'} < 2 m_e$, which is favored by the XENON1T excess~\cite{Boehm:2020ltd}.
Accordingly, the decay into the $e^\pm$ is kinematically forbidden.

As the left-handed charged leptons and the neutrinos are in the same multiplets in the Standard Model (SM), simple introduction of a new $U(1)$ gauge interaction tends to predict $g_{\nu} = g_\ell$. 
It is, however, possible to achieve $g_{\nu} \gg g_\ell$ in, for example, the $U(1)_{L_\mu-L_\tau}$ gauge symmetry~\cite{Foot:1990mn,He:1991qd}. 
In this case, $g_{\nu}$ corresponds to the  $U(1)_{L_\mu-L_\tau}$ gauge coupling, while $g_e$ is provided through the gauge kinetic mixing between the $U(1)_{L_\mu-L_\tau}$ gauge boson and the photon which leads to $g_\nu \simeq 70g_e$~\cite{Holdom:1985ag}.
The dark photon coupling to the SM gauge bosons through the kinetic mixing also predicts hierarchical coupling constants, $g_\nu \sim m_{Z'}^2/m_{Z}^2\times g_e$~\cite{Ibe:2019gpv}.
More generally, the dark photon with non-trivial $U(1)_X$ gauge charge assignment to the SM fermions lead to  gauge couplings, $g_e \sim g_X - \epsilon e$ and $g_\nu \sim g_{X}$, which can be tuned either $g_\nu \gg g_e$ or $g_e \gg g_\nu$.
Here, $e$ and $g_X$ are the QED and $U(1)_X$ gauge coupling constants, and $\epsilon$ is the kinetic mixing parameter between field strengths, i.e., $\epsilon F_{\mathrm{QED}}^{\mu\nu}F_{X\mu\nu}$.%
\footnote{The hierarchical coupling condition ($g_e \gg g_\nu$ or $g_\nu\gg g_e$) is stable against the radiative corrections.
The stability of $g_e \gg g_\nu$ is trivial once we impose $g_\nu = 0$ as the on-shell renormalization condition, because the neutrinos do not couple to the photon nor $Z'$ at any order. 
The stability of $g_\nu\gg g_e$ is clarified by
the conjugation symmetry of QED.
The charge conjugation symmetry forbids the mixing term once we impose $g_e = 0$.
Thus, the radiatively generated mixing term is proportional to $g_e$, and hence, the hierarchical coupling is technically natural.}
In the following,
we take $g_{\nu}$ and $g_e$ to be independent free parameters.
We also assume that the mediator couples to one flavor of the neutrinos.
Since the neutrino oscillation is fast enough for $T< \order{1}$\,MeV, 
the choice of the neutrino flavor is irrelevant for the following arguments.

\begin{figure}[t]
    \begin{minipage}[t]{0.47\linewidth}
    \centering
    \includegraphics[width=0.47\textwidth]{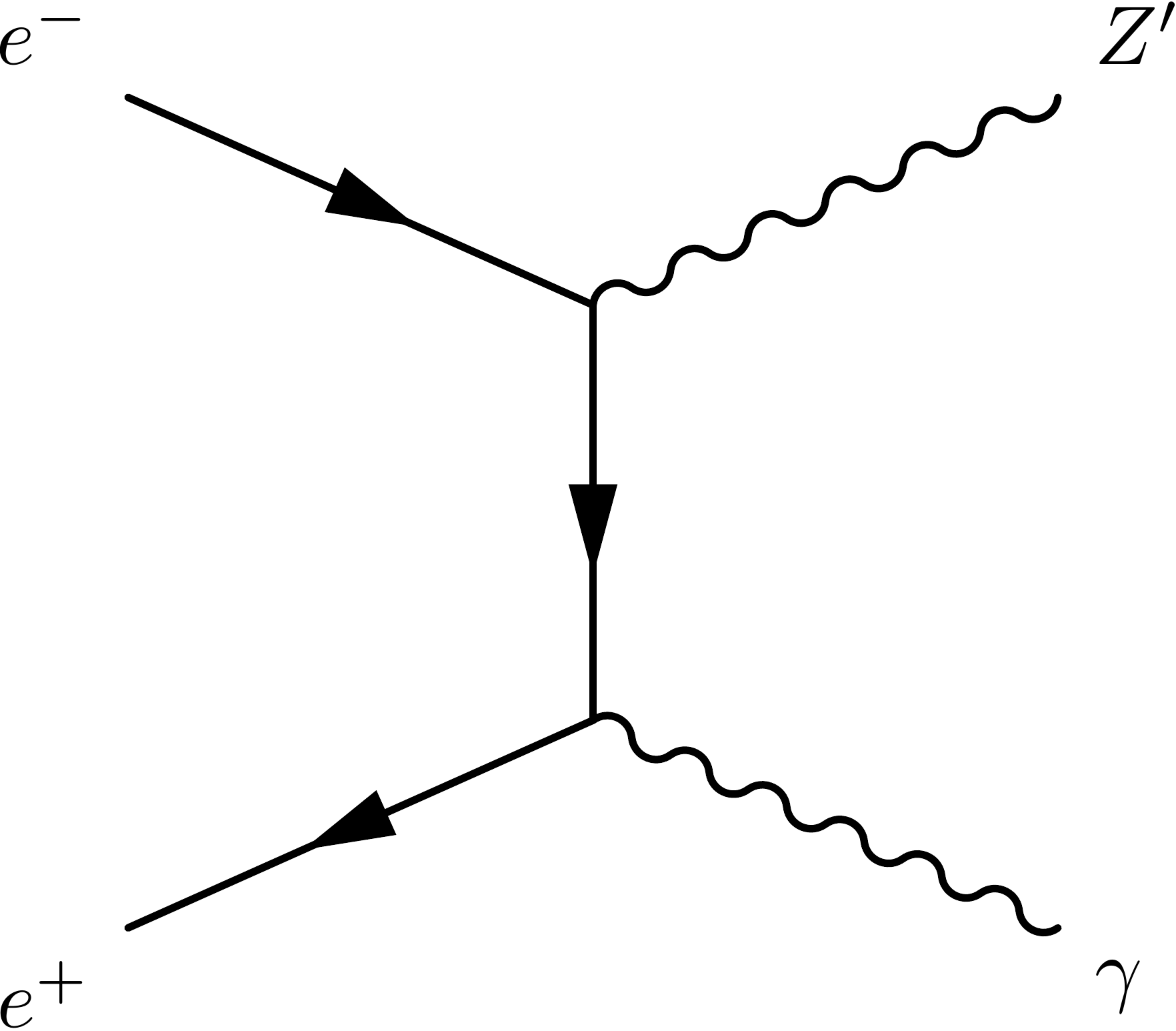}
    \subcaption{electron annihilation}
    \end{minipage}
    \begin{minipage}[t]{0.47\linewidth}
    \centering
    \includegraphics[width=0.47\textwidth]{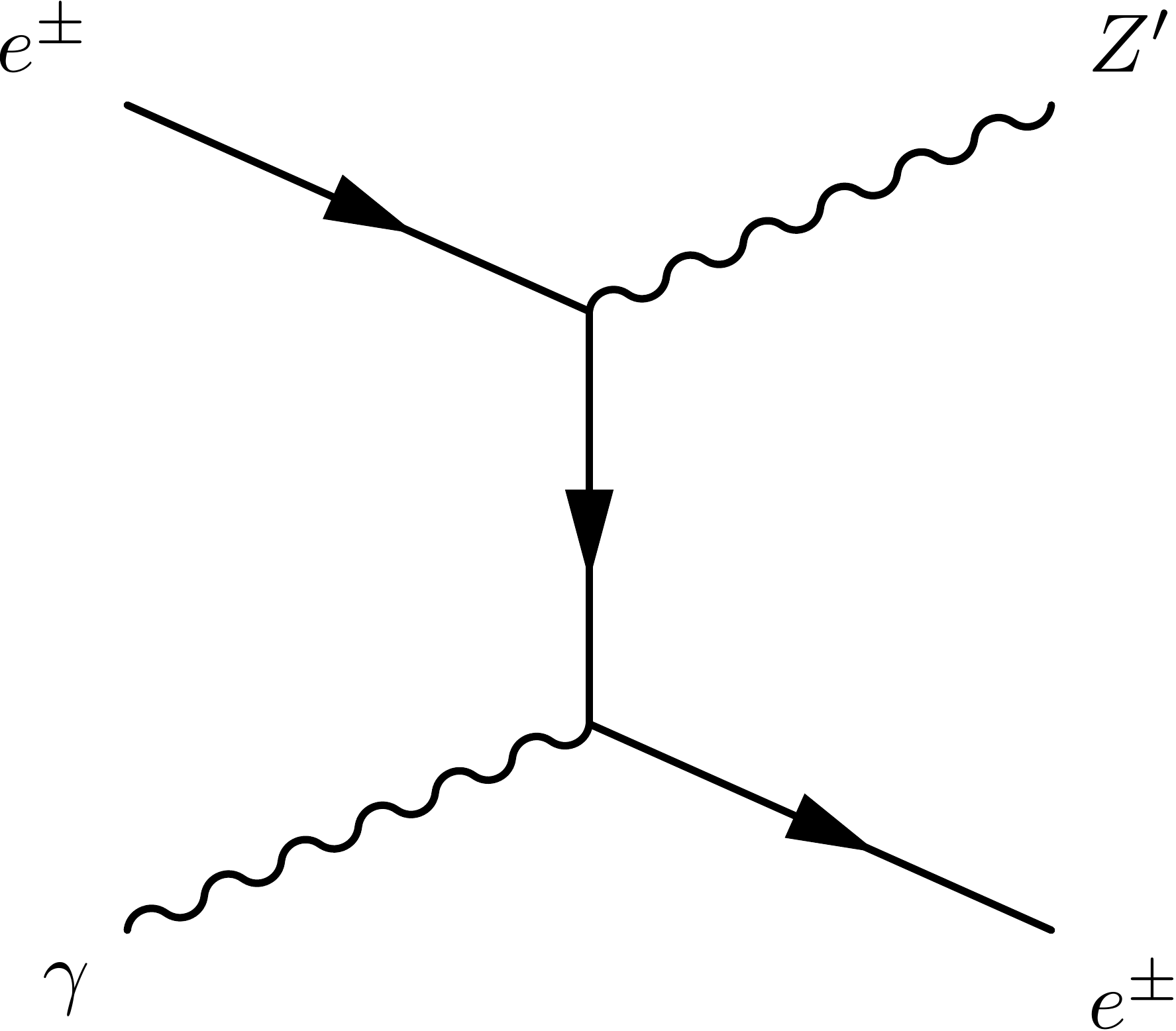}
    \subcaption{electron scattering}
    \end{minipage}\\
    \vspace{1cm}
    \\
	\begin{minipage}[b]{0.47\linewidth}
    \centering
    \includegraphics[width=0.47\textwidth]{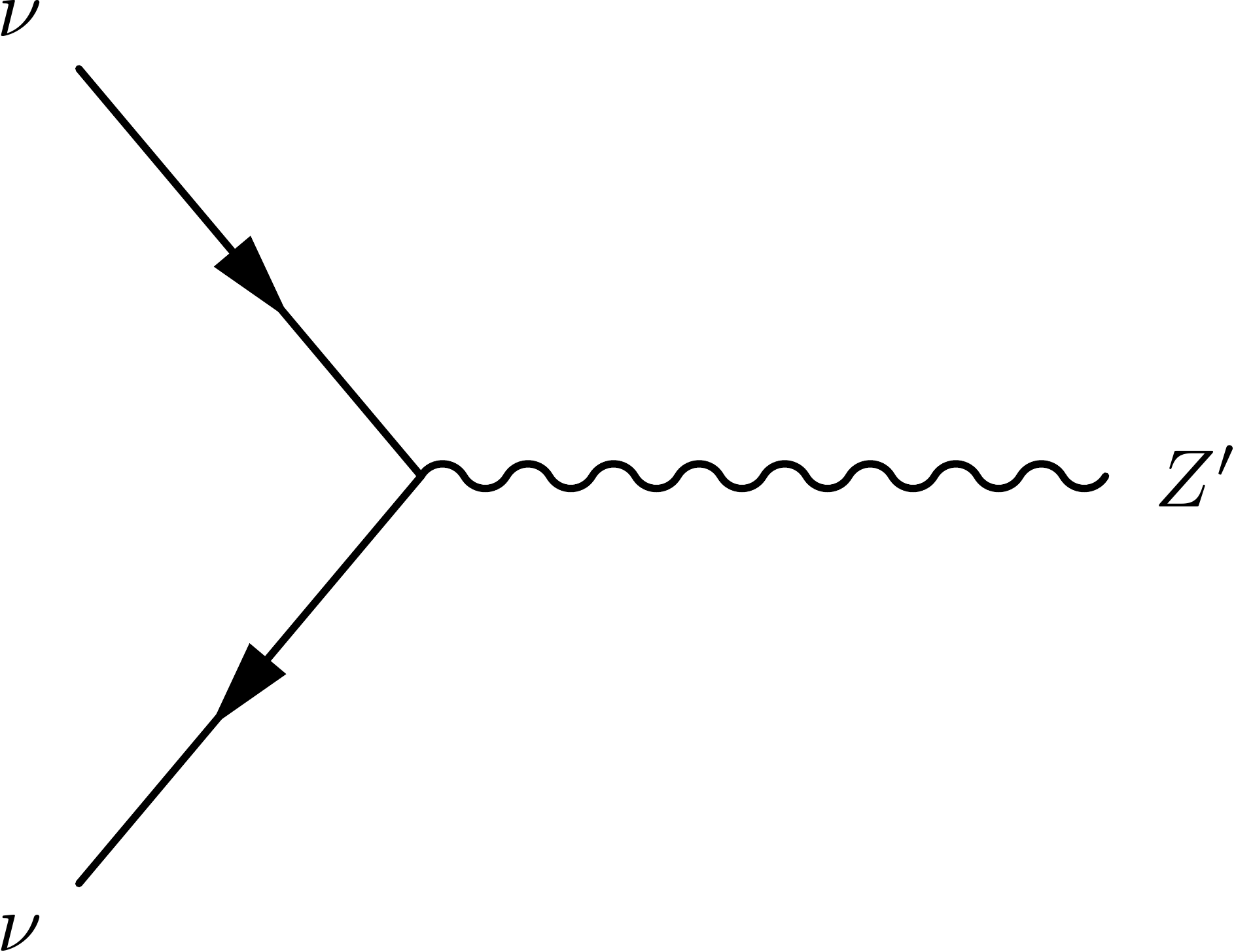}
    \subcaption{neutrino inverse decay}
    \end{minipage}
    \begin{minipage}[b]{0.47\linewidth}
    \centering
    \includegraphics[width=0.47\textwidth]{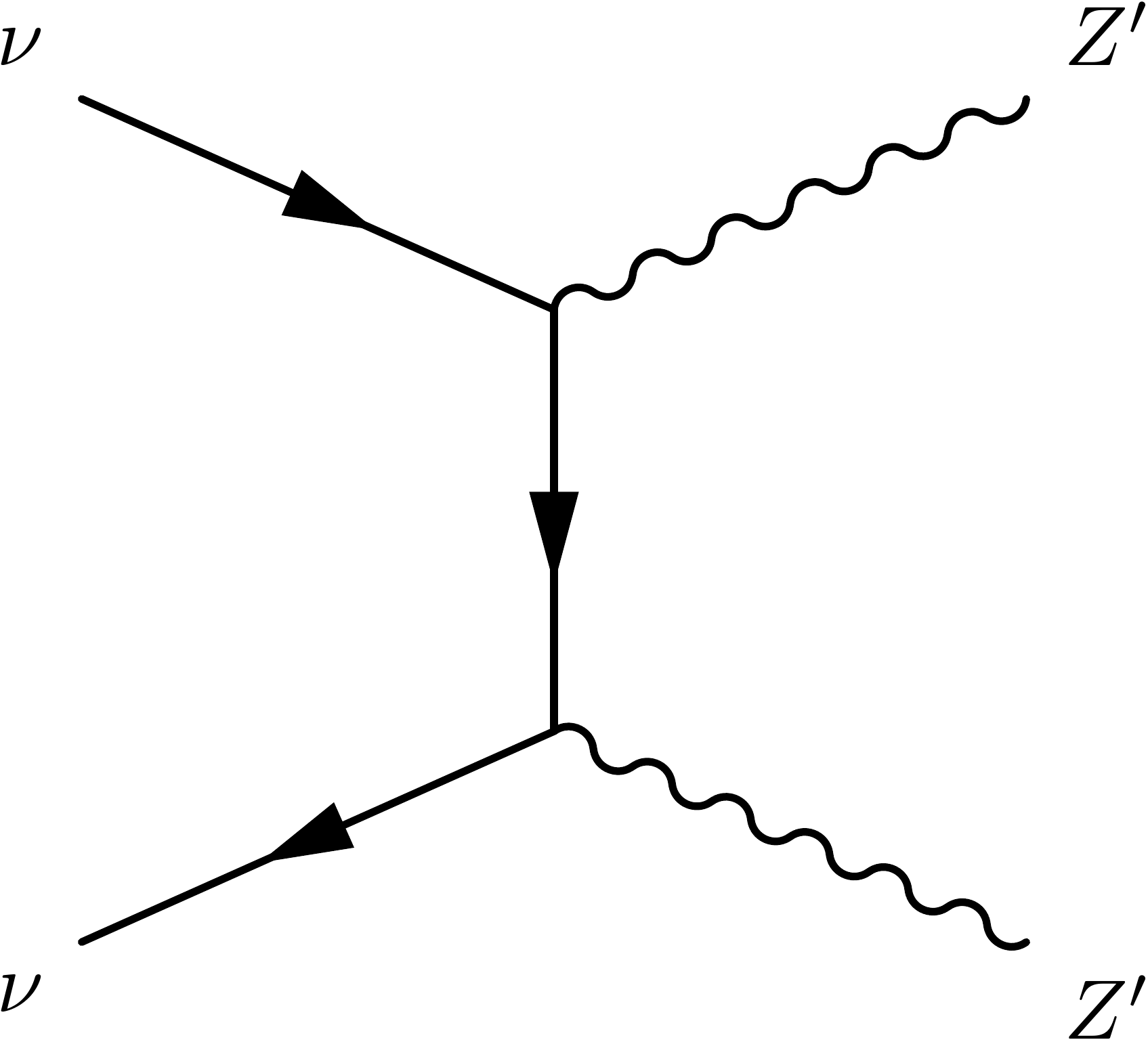}
    \subcaption{neutrino annihilation}
    \end{minipage}
\caption{
The Feynman diagrams relevant for the $Z'$ production.
}\label{fig:diagrams}
\end{figure}

Now, let us discuss cosmology of the mediator at the temperature below $\order{10}$\,MeV, 
which is crucial for the determination of $N_{\mathrm{eff}}$.
In this setup, the mediators are produced from the thermal bath through, $e^-+e^+ \leftrightarrow \gamma + Z'$, $e^\pm + \gamma \leftrightarrow e^\pm  + Z'$,
$\nu+\nu \leftrightarrow  Z'$,
and $\nu+\nu \leftrightarrow  Z' + Z'$ (see Fig.\,\ref{fig:diagrams}).
These production processes are relatively enhanced compared to the Hubble expansion rate as the temperature decreases.
Hence, the mediator is produced at the lower temperature even if it has the zero initial abundance after inflation.
In our analysis, we adopt this ``freeze-in" scenario, which leads to the most conservative constraint. 

In order to comprehend the overview of the cosmology, let us estimate the temperature at which the production processes become effective.
For the electron annihilation or scattering production, if 
\begin{align}
    \Gamma_{e^-e^+\to \gamma Z'}(T)\sim e^2 g_e^2  T \gtrsim H(T)\ ,
\end{align}
that is,
\begin{align}
    T \lesssim e^2g_{e}^2 M_{\mathrm{Pl}}\ ,
\end{align}
these processes are effective.
Here, $H$ is the Hubble expansion rate and $M_{\mathrm{Pl}}$ is the reduced Planck scale.
Thus, the dark photon
is thermalized with
the $\gamma\mbox{-}e$ thermal bath before the $e^\pm$
annihilation for 
\begin{align}
    g_e \gg 10^{-9}\ .
\end{align}
In this region, $Z'$ and $e^\pm$ are thermalized together and share the same temperature.

For the production via the neutrino annihilation, it becomes efficient for
\begin{align}
    \Gamma_{\nu\nu\to Z' Z'}(T)\sim g_\nu^4 T\gtrsim H(T) \quad \therefore T \lesssim g_\nu^4M_{\mathrm{Pl}}\ .
\end{align}
Thus, $Z'$ and $\nu$ are thermalized by the temperature, $T\sim m_{Z'}$, for
\begin{align}
\label{eq:nuscatter}
g_\nu\gg 10^{-6}\left(\frac{m_{Z'}}{1\,\mathrm{keV}}\right)^{1/4}\ .
\end{align}

As we consider $m_{Z'} < 2 m_e$,  the mediator mainly decays into a pair of neutrinos.
The decay rate at the temperature $T\gtrsim m_{Z'}$
is given by
\begin{align}
    &\Gamma_{Z'\to \nu\nu}(T) \sim \Gamma_{Z'\to \nu\nu}^0\times\frac{m_{Z'}}{T}\ .
    \end{align}
Here, $\Gamma^0_{Z'\to \nu\nu}$ denotes the partial decay rate of $Z'$ into one flavor of the three neutrinos at the $Z'$ rest frame,
    \begin{align}
    \label{eq:gamma0}
    &\Gamma_{Z'\to \nu\nu}^0 = \frac{1}{24\pi}g_\nu^2 m_{Z'}\ .
\end{align}
In the following, we assume that the mediator couples to one flavor of the three neutrinos.
We also treat the neutrinos massless throughout this paper.
The light mediator exhibits the in-equilibrium decay at the neutrino temperature $T_\nu$ is of $\order{m_{Z'}}$ for
\begin{align}
\label{eq:inequilibrium}
    g_\nu \gg 5\times 10^{-12} \left(\frac{m_{Z'}}{1\,\mathrm{keV}}\right)^{1/2}\ .
\end{align}
Note that if $g_\nu$ 
is between Eqs.\,\eqref{eq:nuscatter} 
and \eqref{eq:inequilibrium},
only the inverse decay is effective.
In this case, the neutrino thermal bath may have a non-zero chemical potential due to the number conservation of $Z'$ and $\nu$~\cite{Escudero:2019gzq,Escudero:2020dfa}.

When $Z'$ thermalization occurs by the neutrino decoupling era, the energy injected into the $\nu$ sector changes the ratio of the neutrino and the photon energy densities, $\rho_\nu/\rho_\gamma$, from the 
one in the standard cosmology.
In this case, $N_{\mathrm{eff}}$
is changed drastically,
which conflicts with the CMB observations. 

\begin{figure}[t]
\centering
	\includegraphics[height=0.35\vsize,clip]{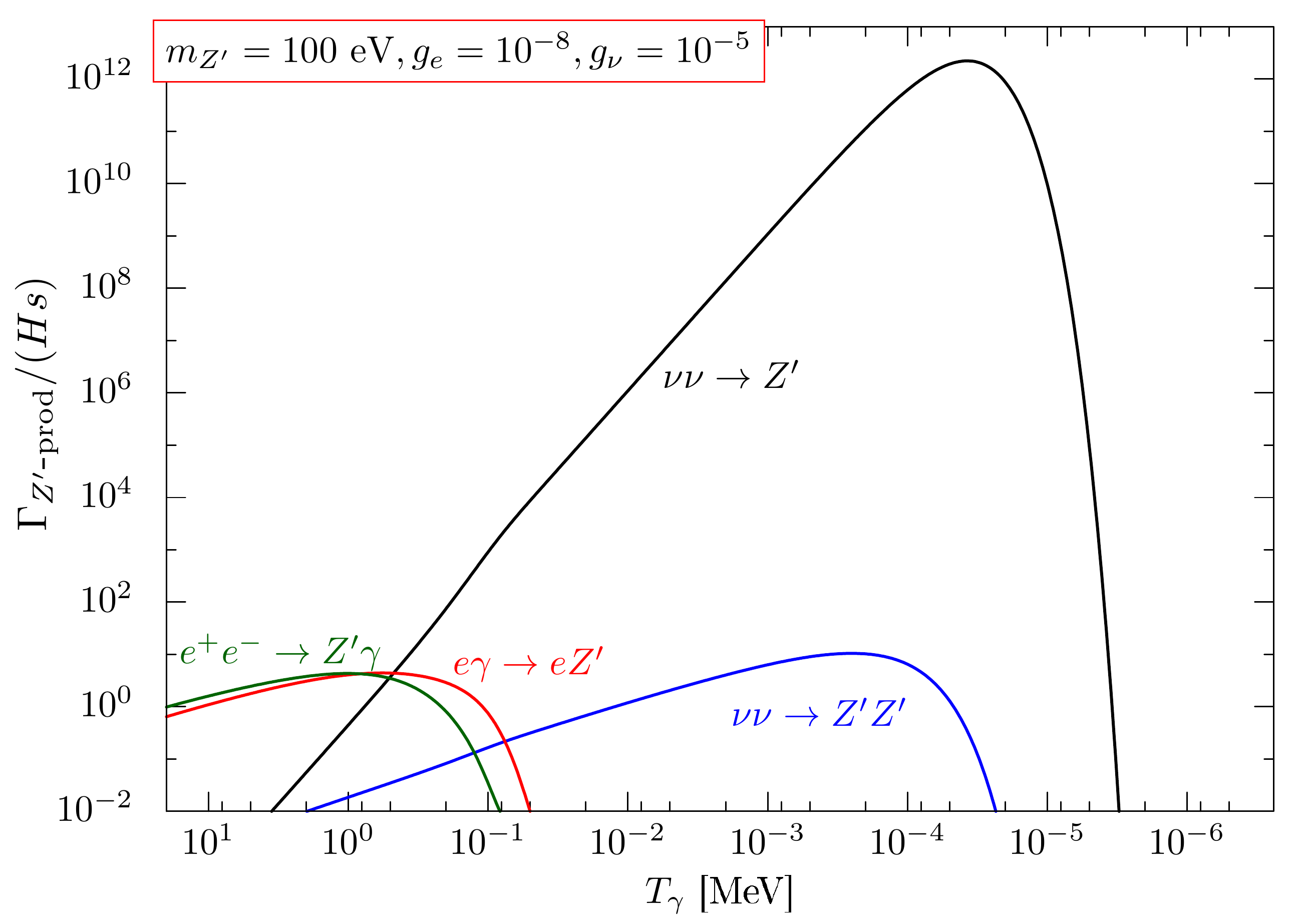}

	\caption{The temperature dependence of the production rates of the modes $e\gamma\to eZ'$ (red), $e^+ e^- \to Z'\gamma$ (green), $\nu\nu\to Z'$ (black) and $\nu\nu\to Z'Z'$ (blue).}
    \label{fig:rate evolution}
\end{figure}

For the summary of this section, we show the temperature dependence of the production rates of $Z'$ in Fig.~\ref{fig:rate evolution}.
Here we define the production rates per unit volume, $\Gamma_{Z'\mbox{-}\mathrm{prod}}$, as
\begin{align}
    \Gamma_{Z'\mbox{-}\mathrm{prod}}= \int \frac{d^3p_{Z'}}{(2\pi)^3}f_{Z'}^{\mathrm{BE}}\Tilde{G}_{Z'},
\end{align}
where $\Tilde{G}_{Z'}$ is a collision term defined in Eqs.~\eqref{eq:Z'decay}, \eqref{eq:Z'annihilation1}, \eqref{eq:Z'annihilation2}, \eqref{eq:Z'scattering1}, and \eqref{eq:Z'scattering2} in the next section.
We also normalize the production rate with $H s$, where $H$ is the Hubble expansion rate and $s$ is the entropy density of the Universe.
If the production rate exceeds the expansion rate, i.e., $\Gamma_{Z'\mbox{-}\mathrm{prod}}/(Hs)\gtrsim 1$, $Z'$ production from the SM thermal bath becomes efficient.

\section{Boltzmann equations}
\label{sec:Boltzmann}

We solve the Boltzmann equation of the phase space distribution of $Z'$, $f_{Z'}$,
the energy densities
of the $\gamma\mbox{-}e$ thermal bath, $\rho_{\gamma e}$, the neutrino thermal bath, $\rho_\nu$ 
and the number density of
the neutrino thermal bath, $n_\nu$.
Here, we need to treat $\rho_{\nu}$ and $n_\nu$ independently, since the neutrino thermal bath obtains a non-vanishing chemical potential from the $Z'$ interaction as we will see later.
Hereafter, we assume that $\gamma$ has the Bose-Einstein distribution, and $e^\pm$ and $\nu$ have the Fermi-Dirac distributions, which gives a good approximation after the neutrino decoupling~\cite{Escudero:2018mvt,Escudero:2019gzq,Escudero:2020dfa}.
We treat the three flavor neutrinos as a fluid with a single temperature/chemical potential to mimic the effect of the neutrino oscillations as in Refs.~\cite{Escudero:2018mvt,Escudero:2019gzq,Escudero:2020dfa}.

The evolution of the momentum distribution for the mediator $Z'$ is determined by the Boltzmann equation,
\begin{align}
\label{eq:Z' Boltzmann}
\pdv{f_{Z'}}{t} - Hp\pdv{f_{Z'}}{p} 
= -\mathcal{C}[f_{Z'}]\ ,
\end{align}
where $H$ is the Hubble expansion rate and $\mathcal{C}[f_{Z'}]$ is a sum of collision terms.
In this work, we include the decay, the scattering and the annihilation processes for the calculation of the collision terms.
With the aid of the formalism in Ref.~\cite{Ibe:2019gpv}, all the $Z'$ collision terms are written in the approximated form
\begin{align}
\label{eq:Z' collision term}
    \mathcal{C}[f_{Z'}]\simeq & \, \Tilde{G}_{Z'\leftrightarrow \nu\nu}(f_{Z'} - f^{\mathrm{BE}}_{Z'}(T_\nu, 2 \mu_\nu))
    +(\Tilde{G}_{Z'Z'\leftrightarrow \nu\nu}f_{Z'} - \Tilde{G}^{\mathrm{eq}}_{Z'Z'\leftrightarrow \nu\nu})\notag \\
    &+\Tilde{G}_{e^\pm Z'\leftrightarrow e^\pm\gamma}(f_{Z'} - f^{\mathrm{BE}}_{Z'}(T_\gamma,0))
    +\Tilde{G}_{\gamma Z'\leftrightarrow e^-e^+}(f_{Z'} - f^{\mathrm{BE}}_{Z'}(T_\gamma,0))\notag \\
    &+  
    \Tilde{G}_{ Z'\leftrightarrow 3\gamma}(
    f_{Z'}(E_{Z'}) - f^{\mathrm{BE}}_{Z'}(T_\gamma,0)
    )\ .
\end{align}
Here, $f^{\mathrm{BE}}_{Z'}$ is the Bose-Einstein distribution function for a single degree of freedom of $Z'$, $\mu_\nu$ is a chemical potential of the neutrino,  and $T_\gamma, T_\nu$ are
the temperatures 
of the $\gamma\mbox{-}e$ and the neutrino thermal bathes, 
respectively.
Each term for the process including neutrinos is written as
\begin{align}
\label{eq:Z'decay}
    \Tilde{G}_{Z'\leftrightarrow \nu \nu}&=
    \frac{m_{Z'}\Gamma_{Z'\to \nu\nu}(1+\varphi(T_\nu,\mu_\nu,p_{Z'}))}{E_{Z'}}\ ,\\
     \Tilde{G}_{Z' Z'\leftrightarrow \nu\nu}&=
     \frac{1}{512\pi^3} \frac{ 1}{|\ps_{Z'}|E_{Z'}}\int d\Tilde{E}_{Z'}[f_{Z'}(\Tilde{E}_{Z'})-(1+f_{Z'}(\Tilde{E}_{Z'}))e^{-(E_{Z'}+\Tilde{E}_{Z'})/T_\nu}e^{2\mu_\nu/T_\nu}]\notag \\
     \label{eq:Z'annihilation1}
    &\hspace{5cm}\times\int  ds  \frac{1}{\sqrt{s} |\ps^{\mathrm{cms}}_{Z'Z'}|}
 \int dt |\bar{\mathcal{M}}_{Z'Z'\leftrightarrow \nu\nu}|^2
    \times 12\ , \\
    \Tilde{G}^{\mathrm{eq}}_{Z' Z'\leftrightarrow \nu\nu}&=
    \frac{1}{512\pi^3} \frac{ 1}{|\ps_{Z'}|E_{Z'}}\int d\Tilde{E}_{Z'}[(1+f_{Z'}(\Tilde{E}_{Z'}))e^{-(E_{Z'}+\Tilde{E}_{Z'})/T_\nu}e^{2\mu_\nu/T_\nu}]\notag \\
    \label{eq:Z'annihilation2}
    &\hspace{5cm}\times\int  ds  \frac{1}{\sqrt{s} |\ps^{\mathrm{cms}}_{Z'Z'}|}
 \int dt |\bar{\mathcal{M}}_{Z'Z'\leftrightarrow \nu\nu}|^2
    \times 12\ .
\end{align}
Here, we define 
\begin{align}
\label{eq:phi}
    \varphi(T_\nu,\mu_\nu,p_{Z'}) = \frac{m_{Z'} T_\nu}{p_{Z'} p^0_\nu}
    \log\left( 
    \frac{e^{(E_{Z'} E^0_\nu-\mu_\nu m_{Z'})/(T_\nu m_{Z'})} + e^{ -p_{Z'} p^0_\nu/(T_\nu m_{Z'}) }}{
    e^{(E_{Z'} E^0_\nu - \mu_\nu m_{Z'})/(T_\nu m_{Z'})} + e^{ p_{Z'} p^0_\nu/(T_\nu m_{Z'}) }
        }
    \right)\ ,
\end{align}
where $p^0_\nu = E^0_\nu = m_{Z'}/2$ are the momentum and energy of the neutrino at the rest frame of the dark photon, $Z'$.
For collision terms of the $\nu\nu$ annihilation processes, we use the integrated amplitude
\begin{align}
    &\int  ds  \frac{1}{\sqrt{s} |\ps^{\mathrm{cms}}_{Z'Z'}|}
 \int dt |\bar{\mathcal{M}}_{Z'Z'\leftrightarrow \nu\nu}|^2 = -\frac{8}{9}g_{\nu}^4 m_{Z'}^2\qty[(x_{+}-x_{-}) +2 (f(y_{+})-f(y_{-}))]\ ,\\
  &f(y)=\frac{1}{2y^2}+\frac{y^2}{2}+\frac{\log(y)}{y^2}-y^2\log(y)-4\log(y)^2+4\int^\infty_{y^2} dz \frac{\log(z)}{z^2+1}\ ,\notag \\
  &x_\pm = \qty(\frac{E_{Z'}}{m_{Z'}}+\frac{\Tilde{E}_{Z'}}{m_{Z'}})^2\pm \sqrt{\left(\frac{E_{Z'}}{m_{Z'}}\right)^2-1}\sqrt{\left(\frac{\Tilde{E}_{Z'}}{m_{Z'}}\right)^2-1}\notag\ , \\
  &y_\pm = \frac{1}{2}\qty(\sqrt{x_\pm}+\sqrt{x_\pm -4})\ .\notag
\end{align}
For the scattering process with the $\gamma\mbox{-}e$ thermal bath, we use the same formulae in the appendix of Ref.~\cite{Ibe:2019gpv} with parameters replaced as $\varepsilon g\to g_e, m_{\gamma'}\to m_{Z'}$,
\begin{align}
\label{eq:Z'scattering1}
    \Tilde{G}_{e^\pm Z'\leftrightarrow e^\pm\gamma}&= \frac{1}{512\pi^3} \frac{T e^{-E_{Z'}/T}}{|\ps_{Z'}|E_{Z'}f^{\mathrm{BE}}_{Z'}(E_{Z'})}\int  ds  \frac{1}{\sqrt{s} |\ps^{\mathrm{cms}}_{eZ'}|}
    \log\left[\frac{1+e^{-E_e^-/T}}{1+e^{-E_e^+/T}}\right]
 \int dt |\bar{\mathcal{M}}_{e^\pm\gamma\leftrightarrow e^\pm Z'}|^2
    \times 8\ ,\\
    \label{eq:Z'scattering2}
    \Tilde{G}_{\gamma Z'\leftrightarrow e^-e^+}&=\frac{1}{512\pi^3} \frac{T e^{-E_{Z'}/T}}{|\ps_{Z'}|E_{Z'}f^{\mathrm{BE}}_{Z'}(E_{Z'})}\int  ds  \frac{1}{\sqrt{s} |\ps^{\mathrm{cms}}_{\gamma Z'}|}
 \log\left[\frac{1-e^{-E_\gamma^+/T}}{1-e^{-E_\gamma^-/T}}\right]
    \int dt |\bar{\mathcal{M}}_{e^+e^-\leftrightarrow\gamma Z'}|^2\times 8\ ,\\
    \Tilde{G}_{Z'\leftrightarrow 3\gamma}&=\frac{m_{Z'}}{E_{Z'}}\Gamma_{Z'\to 3\gamma} = \frac{17g_e^2 \alpha^3}{2^93^65^3\pi^4}\frac{m_{Z'}^9}{m_e^8}\mathcal{F}(m_{Z'}^2/m_e^2)
    \ .
\end{align}
In the above, we define $|\bar{\mathcal{M}}|^2$ as the amplitude squared averaged over spins of all the initial and final states.
Thus, we multiply factors of spin degrees of freedom when we integrate $|\bar{\mathcal{M}}|^2$ over phase space volumes.
The factor $\mathcal{F}(x)$ in the 3-$\gamma$ decay process is given in Ref.~\cite{McDermott_2018}.

We determine the thermal evolution of the SM particles by solving the zeroth and first moment of the Boltzmann equations,

\begin{align}
\label{eq:n_nu Boltzmann}
         \dv{n_\nu}{t} = -&3Hn_{\nu} 
         - C^{(0)}_{e\leftrightarrow\nu}(T_{\gamma}, T_{\nu}, \mu_{\nu})
         + 2C^{(0)}_{Z'\rightarrow\nu_e\bar{\nu}_e}(T_{\nu}, \mu_{\nu}) 
         + C^{(0)}_{Z'Z'\rightarrow\nu_e\bar{\nu}_e}(T_{\nu}, \mu_{\nu}) \ , \\
         \label{eq:rho_nu Boltzmann}
         \dv{\rho_\nu}{t} = -&4H\rho_{\nu} 
         - C^{(1)}_{e\leftrightarrow\nu}(T_{\gamma}, T_{\nu}, \mu_{\nu})
         + C^{(1)}_{Z'\rightarrow\nu_e\bar{\nu}_e}(T_{\nu}, \mu_{\nu}) 
         + C^{(1)}_{Z'Z'\rightarrow\nu_e\bar{\nu}_e}(T_{\nu}, \mu_{\nu})\ , \\
         \label{eq:rho_gamma Boltzmann}
         \dv{\rho_{\gamma e}}{t} = -&3H(\rho_{\gamma e} + p_{\gamma e}) 
         + C^{(1)}_{e\leftrightarrow\nu}(T_{\gamma}, T_{\nu}, \mu_{\nu})
         + C^{(1)}_{eZ'\leftrightarrow e\gamma}(T_{\gamma}) 
         + C^{(1)}_{\gamma Z'\leftrightarrow e^-e^+}(T_{\gamma})
         + C^{(1)}_{ Z'\leftrightarrow 3\gamma}(T_{\gamma})\ ,
\end{align}
where $C^{(n)}$
is the $n$-th energy moment of a collision term.
Here, $n_{\nu} = n_{\nu_e} + n_{\nu_{\mu}} + n_{\nu_{\tau}}$ and $\rho_{\nu} = \rho_{\nu_e} + \rho_{\nu_{\mu}} + \rho_{\nu_{\tau}}$ are 
the number and energy density of the neutrinos, and $\rho_{\gamma e}$ and $p_{\gamma e}$ are the density and pressure 
of the $\gamma\mbox{-}e$ thermal bath, which include thermal corrections~\cite{Heckler:1994tv,Fornengo:1997wa,Mangano:2001iu}. 
For the processes including $Z'$, we integrate the each term of the right hand side of Eq.\eqref{eq:Z' collision term} to obtain $C^{(n)}$.
Following the procedure of Ref.\,\cite{Escudero:2020dfa}, we have included the effect of spin-statistics and the electron mass
in the collision terms between $e^\pm$ and the neutrinos, $C^{(0)}_{e\leftrightarrow\nu}$ and $C^{(1)}_{e\leftrightarrow\nu}$, which are  summarized in the Appendix~\ref{sec:neutrino collision}.

The neutrino momentum distributions 
depart from the Fermi-Dirac 
distributions
at low 
temperatures unless the reaction $Z\leftrightarrow \nu$  is thermalized~\cite{deSalas:2016ztq,Escudero:2019gzq}.
The above approximation is validated in Ref.\,\cite{Escudero:2020dfa} by comparing with the full solution of the Boltzmann equation of the neutrino distributions, which shows the accuracy is better than $0.04$\%.

The new interactions in $\mathcal{L}_{Z'}$ also induces a new channel of $\nu\mbox{-}e$ scattering via the off-shell $Z'$ exchange.
As shown in Ref.\,\cite{Boehm:2020ltd}, however, rather small couplings $\sqrt{g_\nu g_e}\sim 3\times 10^{-7}$ are preferred to explain the XENON1T excess.
In such a small coupling region, the $\nu\mbox{-}e$ scattering mediated by the off-shell $Z'$ is negligible compared to the weak interactions at around the neutrino decoupling temperature, $T_{\nu\mbox{-}\mathrm{dec}}=\order{1}\,\mathrm{MeV}$.
Thus, we ignore the $Z'$ mediated scattering process.

We set the following initial conditions of the Boltzmann equations
at $T_{\rm init} = T_{\gamma e} = T_{\nu} = 20$ MeV,
\begin{align}
& f_{Z'} = 0\ ,  \\
& \rho_{\gamma e} = \rho_{\gamma}^{\mathrm{BE}}(T_{\mathrm{init}})+
\rho_{e}^{\mathrm{FD}}(T_{\mathrm{init}}) +  \rho_{\gamma e}^{\mathrm{QED}}(T_{\mathrm{init}}) \ , \\
&\rho_\nu = \rho_\nu^{\mathrm{FD}}(T_{\mathrm{init}},0) \ , \\ 
&n_\nu = n_\nu^{\mathrm{FD}}(T_{\mathrm{init}},0)\ ,
\end{align}
where $\rho_{\gamma e}^{\mathrm{QED}}$ is the QED loop correction to the
electromagnetic energy density.
We solve the Boltzmann equations,
\eqref{eq:Z' Boltzmann}, \eqref{eq:n_nu Boltzmann}, \eqref{eq:rho_nu Boltzmann} and \eqref{eq:rho_gamma Boltzmann} numerically
with binned $f_{Z'}$.
We validated our numerical code by comparing with the thermalized limit discussed in the next section.

\section{\texorpdfstring{$N_{\mathrm{eff}}$}{Lg} constraint}
\label{sec:Neffconstraint}
\begin{figure}[H]
	\centering
	{\includegraphics[width=0.47\textwidth]{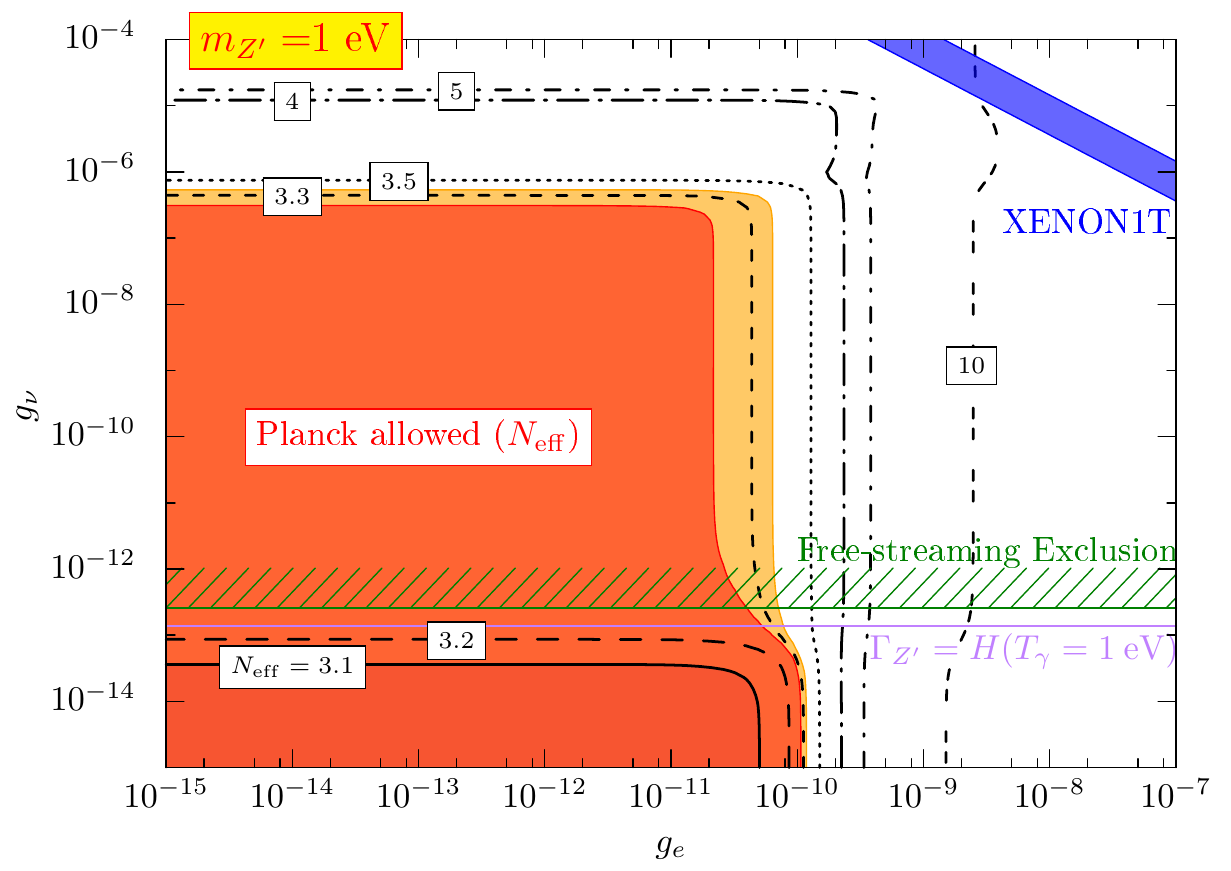}}
	{\includegraphics[width=0.47\textwidth]{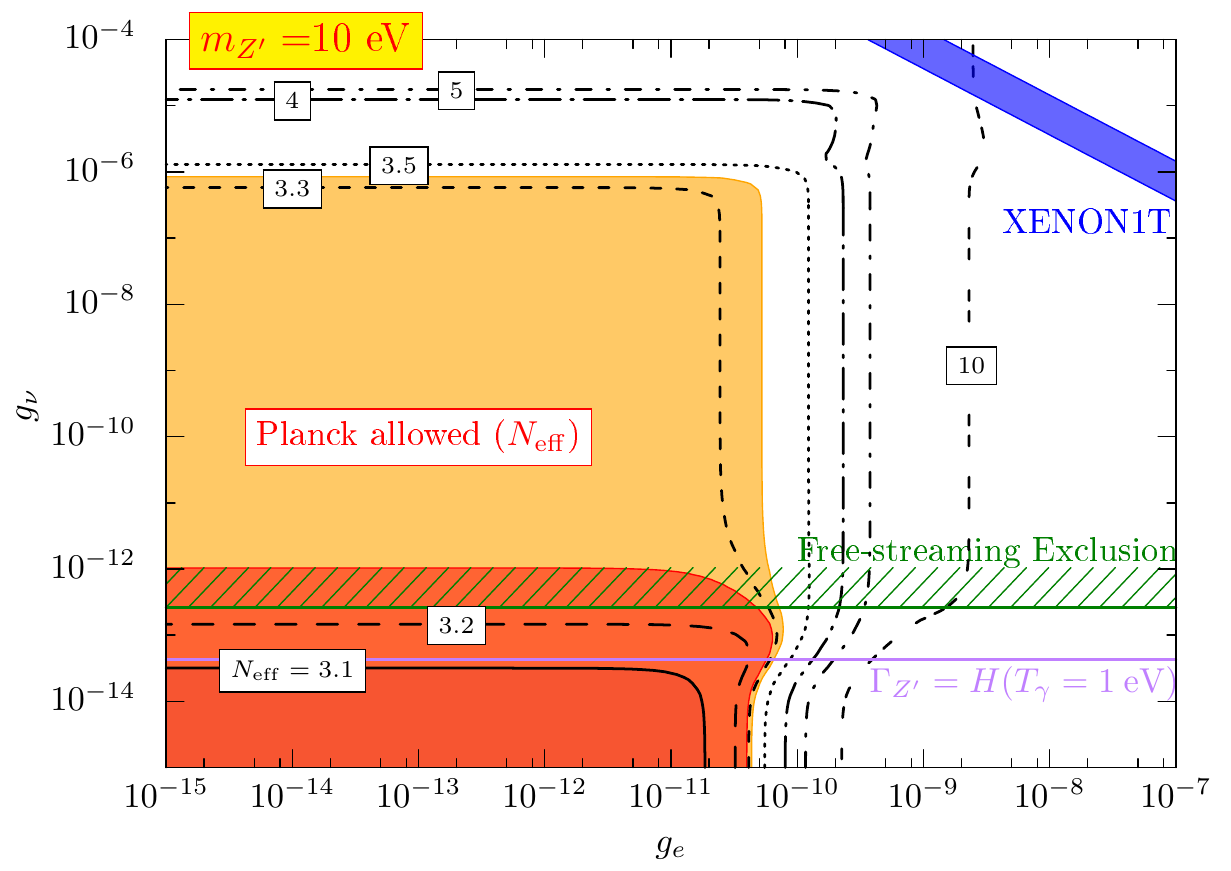}}
	
    \vspace{.4cm}

	{\includegraphics[width=0.47\textwidth]{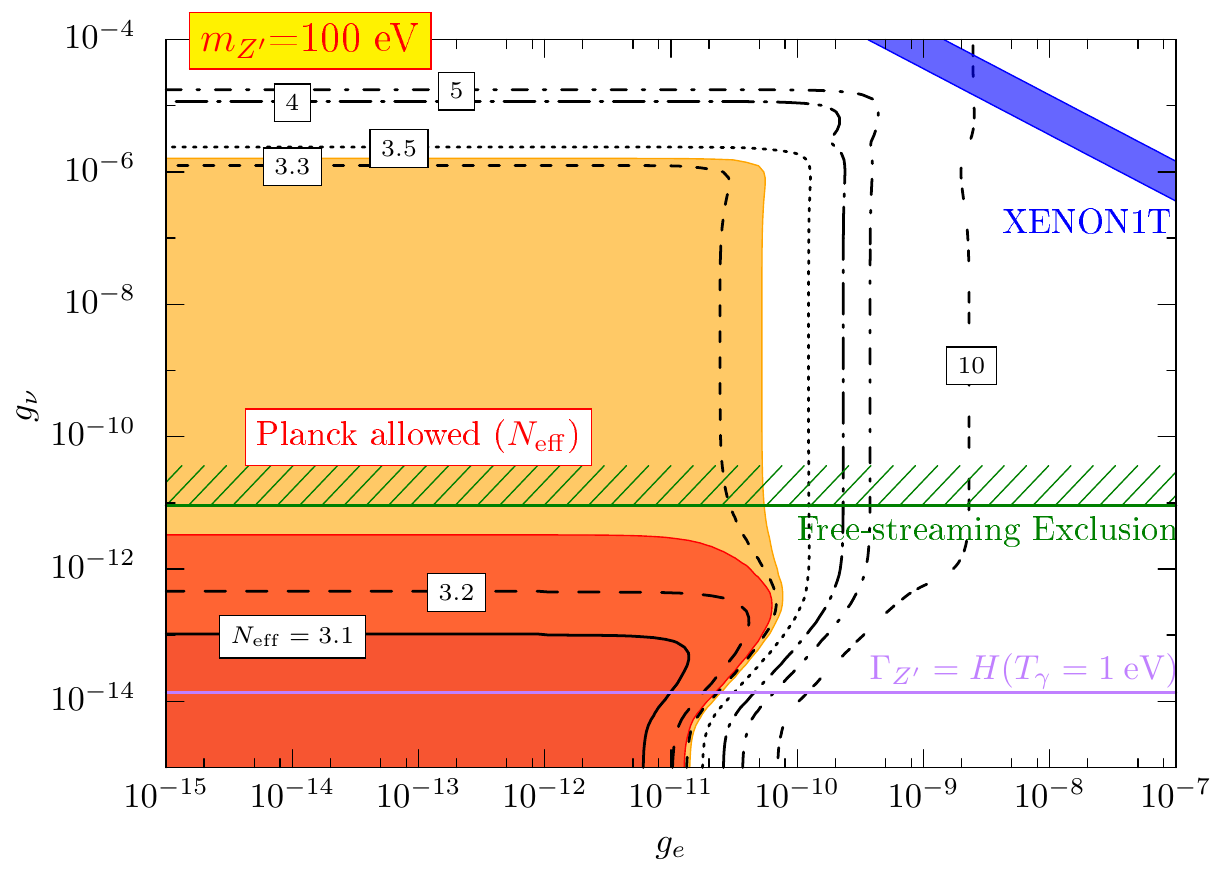}}
	{\includegraphics[width=0.47\textwidth]{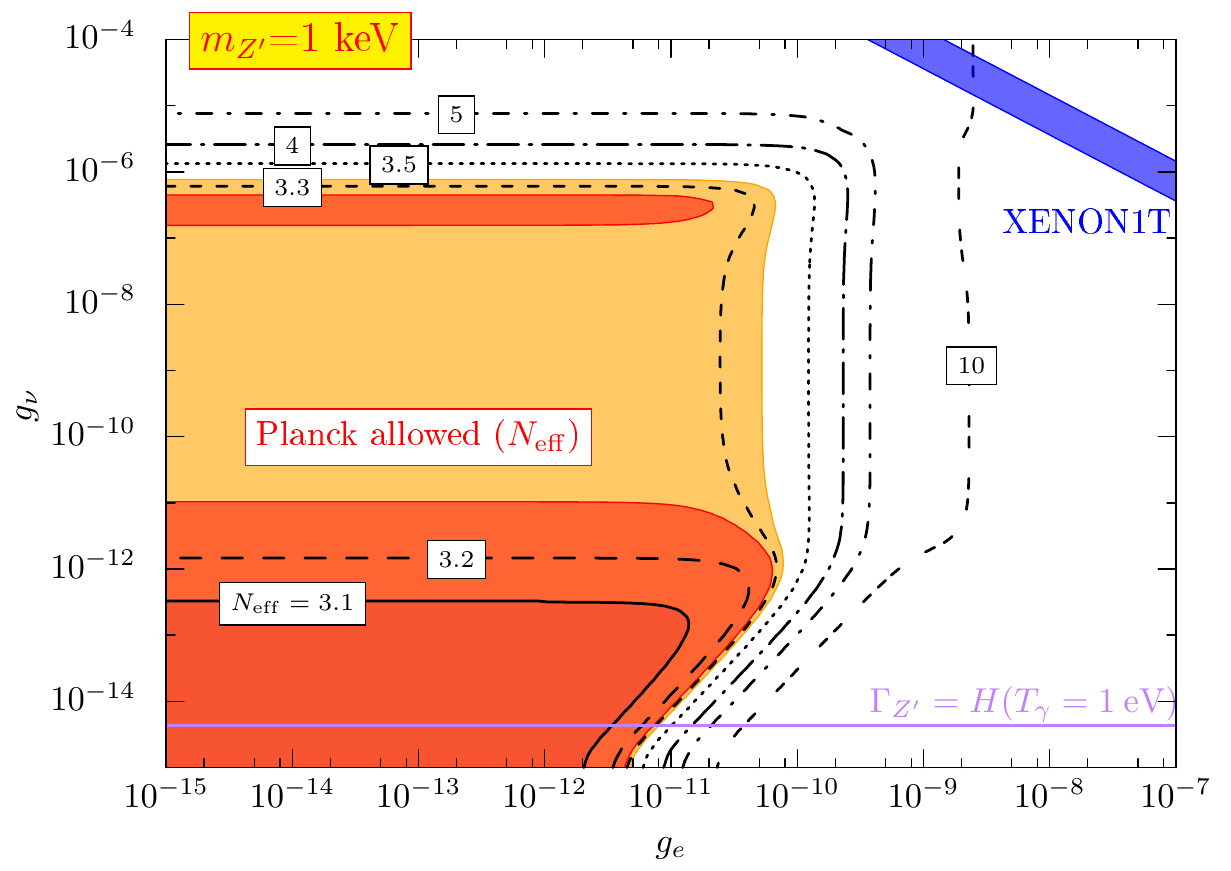}}
		
    \vspace{.4cm}
	
	{\includegraphics[width=0.47\textwidth]{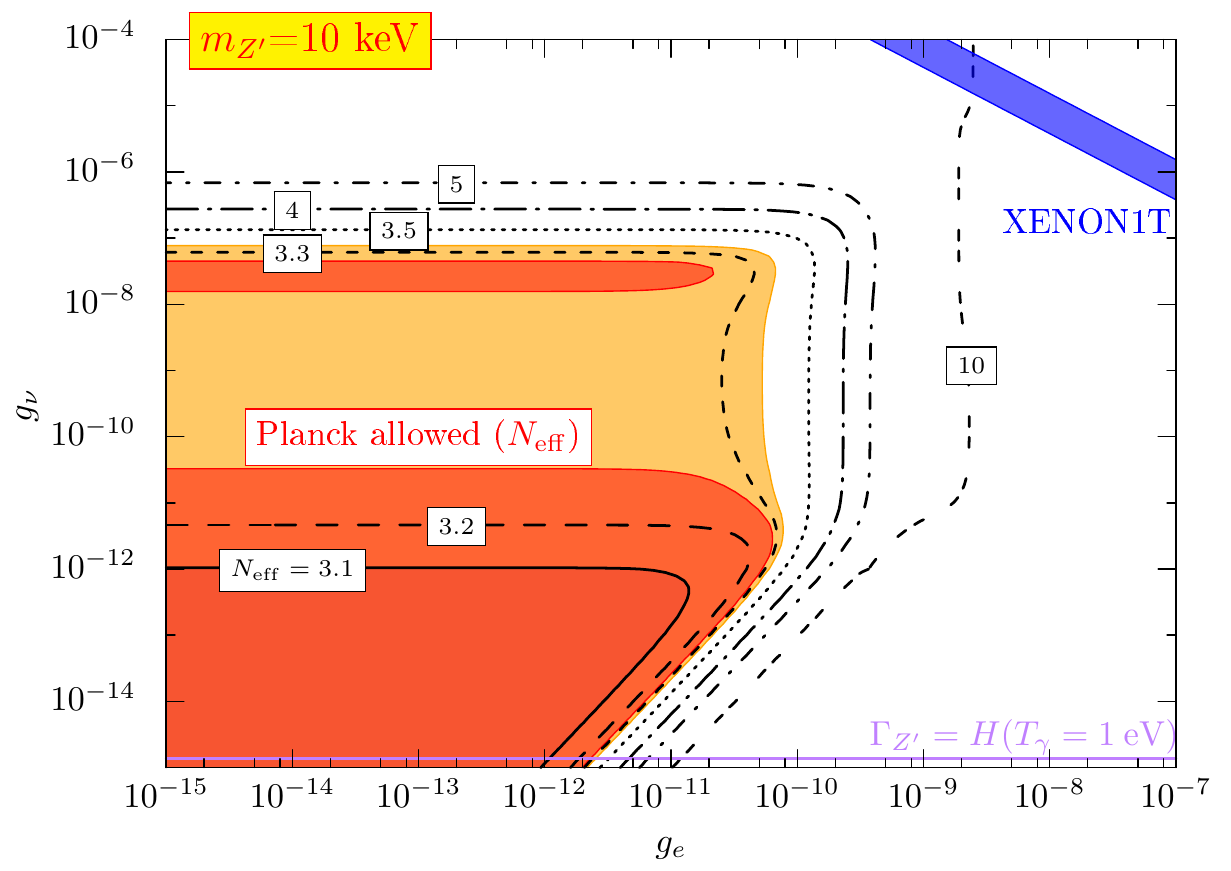}}
	{\includegraphics[width=0.47\textwidth]{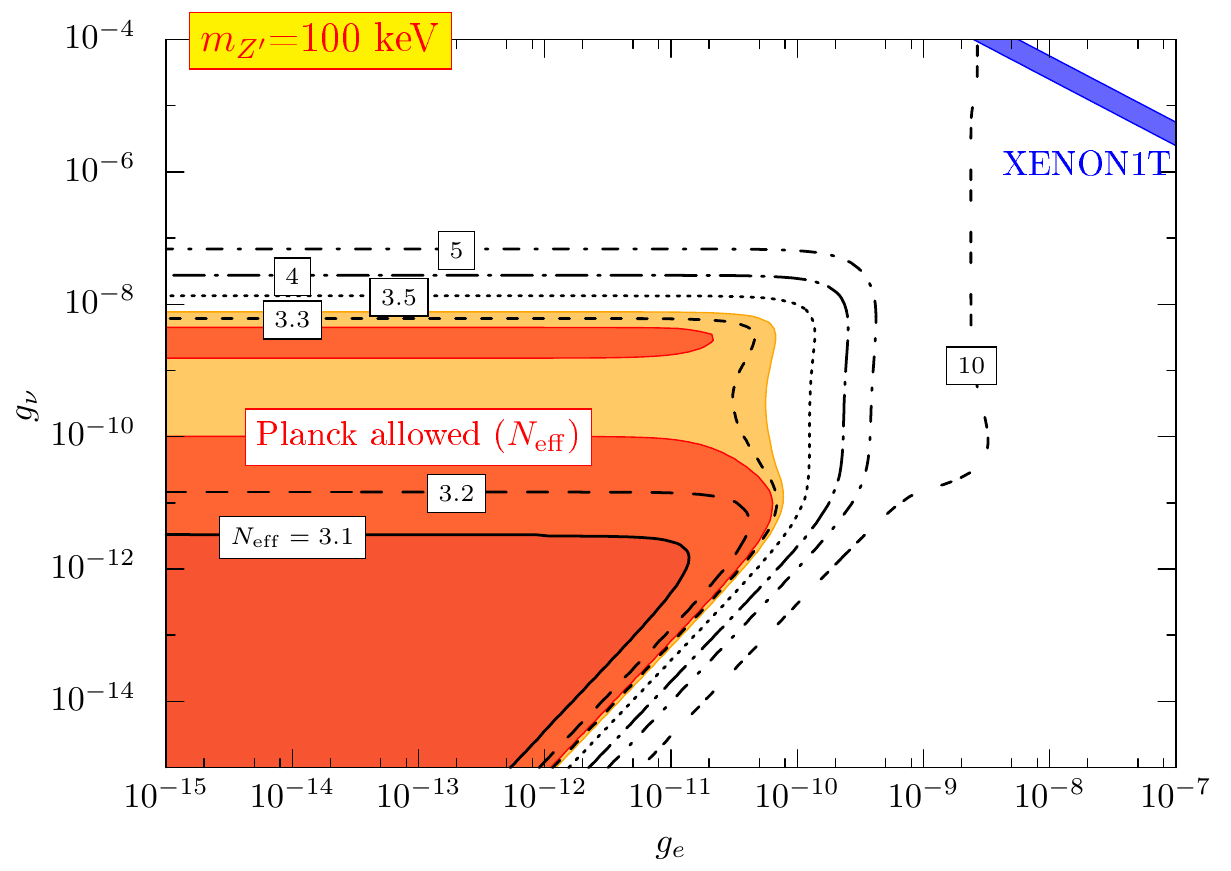}}
\caption{
The contour plot of $N_{\rm eff}$
on the ($g_e$, $g_\nu$) plane for a given mediator mass.
Here, we assume that the mediator couples to one flavor of the three neutrinos.
The dark-orange (light-orange) shaded regions are consistent with the Planck CMB only (joint Planck+BAO) constraint, $N_{\rm eff} = 2.92^{+0.36}_{-0.37}$  ($ 2.99^{+0.34}_{-0.33}$) at 95\%\,C.L.
The green line corresponds to the neutrino free-streaming bound~\cite{Escudero:2019gvw}.
The blue bands are the regions favored by the XENON1T excess.
The horizontal purple lines show the parameter at which the total decay rate of the $Z'$ boson is compatible with the Hubble rate at the  CMB era in the standard cosmology.
}\label{fig:Neff}
\end{figure}

Here, we show the results for
the freeze-in scenario of $Z'$.
Fig.\,\ref{fig:Neff} shows the contour plots of $N_{\mathrm{eff}}$
on the $(g_{e}, g_{\nu})$ plane for
$m_{Z'}= 1$\,eV, $10$\,eV, $100$\,eV, $1$\,keV, $10$\,keV, and  $100$\,keV.
Here, $N_\mathrm{eff}$ is defined by,
\begin{align}
\label{eq:Neff}
    N_{\mathrm{eff}} = \frac{8}{7}\left(\frac{11}{4}\right)^{4/3}
\frac{\rho_{\nu}+ \rho_{Z'}}{\rho_{\gamma}}\ ,
\end{align}
at $T_{\gamma} = 0.26$\,eV.
In the figure, we show the contours of $N_{\mathrm{eff}}\le 10$.
In each plot, the dark-orange (light-orange) shaded 
region shows the consistent region with the Planck CMB only (joint Planck+BAO) 
constraint, $N_{\rm eff} = 2.92^{+0.36}_{-0.37}$  ($ 2.99^{+0.34}_{-0.33}$) at 95\% C.L.~\cite{Aghanim:2018eyx}.
The each blue band is the parameter region favored for the light vector mediator interpretation of the XENON1T excess~\cite{Boehm:2020ltd},  i.e., $\sqrt{g_e g_\nu} \sim 3\times 10^{-7}$.%
\footnote{In Ref.~\cite{Boehm:2020ltd}, the mediator universally couples to the three flavors of the neutrinos. Since our constraints are on the mediator coupling to one flavor neutrino,
we scale the $(g_\nu g_e)^{1/2}$ in the favored region by a factor $\sim \sqrt{2}$.}
The figure shows that $N_\mathrm{eff}$ in the favored region  exceeds $N_\mathrm{eff} = 5$
for $m_{Z'} \ge 1$\,eV.

It should be noted that the massive mediator becomes long-lived enough to behave like  ``dark matter" at the recombination time
in the region below the purple dashed lines.
In this case, the above $N_{\mathrm{eff}}$ constraint cannot be applied.
However, since such a region is far off from the favored region for the XENON1T excess, our conclusions are not affected.

In addition to the $N_{\mathrm{eff}}$ constraint, there is also a constraint on $Z'$ with a mass $m_{Z'}\lesssim 100$\,eV from neutrino free-streaming, due to the on-shell production of $\nu\nu\to Z'$.
The mean free path of $\nu$ is affected by the partial decay rate of $Z'$ into the neutrinos~\cite{Chacko:2003dt, Escudero:2019gvw}.
We show the 95\% C.L. bound of this effect in the Fig.~\ref{fig:Neff} with the green shaded line.
We mapped the constraint of Ref.~\cite{Escudero:2019gvw} by matching the Majoron $\phi$ partial decay rate and $Z'$ decay rate \eqref{eq:gamma0}.
\footnote{In this conversion, we take the difference of the degrees of freedom of $Z'$ and the Majoron $\phi$ into account, by multiplying a factor $3$, and average over neutrino flavors due to the neutrino oscillation, by dividing by another factor $3$.

\begin{align}
    \frac{\lambda^2}{16\pi}m_\phi=\Gamma_{Z'\to\nu_e\nu_e}^0\, ,
\end{align}
where $m_\phi$ is a Majoron mass and $\lambda$ is a Majoron-neutrino coupling constant.
}

As we discussed in Sec.\,\ref{sec:model}, the mediator production through $e^\pm+\gamma  \leftrightarrow e^\pm + Z'$
and $e^+ +  e^- \leftrightarrow 
\gamma + Z'$ become less effective for $g_e\ll 10^{-10}$.
The mediator production through $\nu + \nu \leftrightarrow 
Z'+ Z'$ also becomes less effective for $g_\nu \ll 10^{-5}$.
On the other hand, the mediator production via the inverse decay remains effective as long as $g_\nu$ satisfies the inequality in Eq.\,\eqref{eq:inequilibrium}.
Let us summarize the expected value of $N_\mathrm{eff}$ when $Z'$ is thermalized
for various parameter regions.

\newpage

\begin{figure}[H]
	\centering
 	\subcaptionbox{Case (i) \label{fig:case1}}
	{\includegraphics[width=0.47\textwidth]{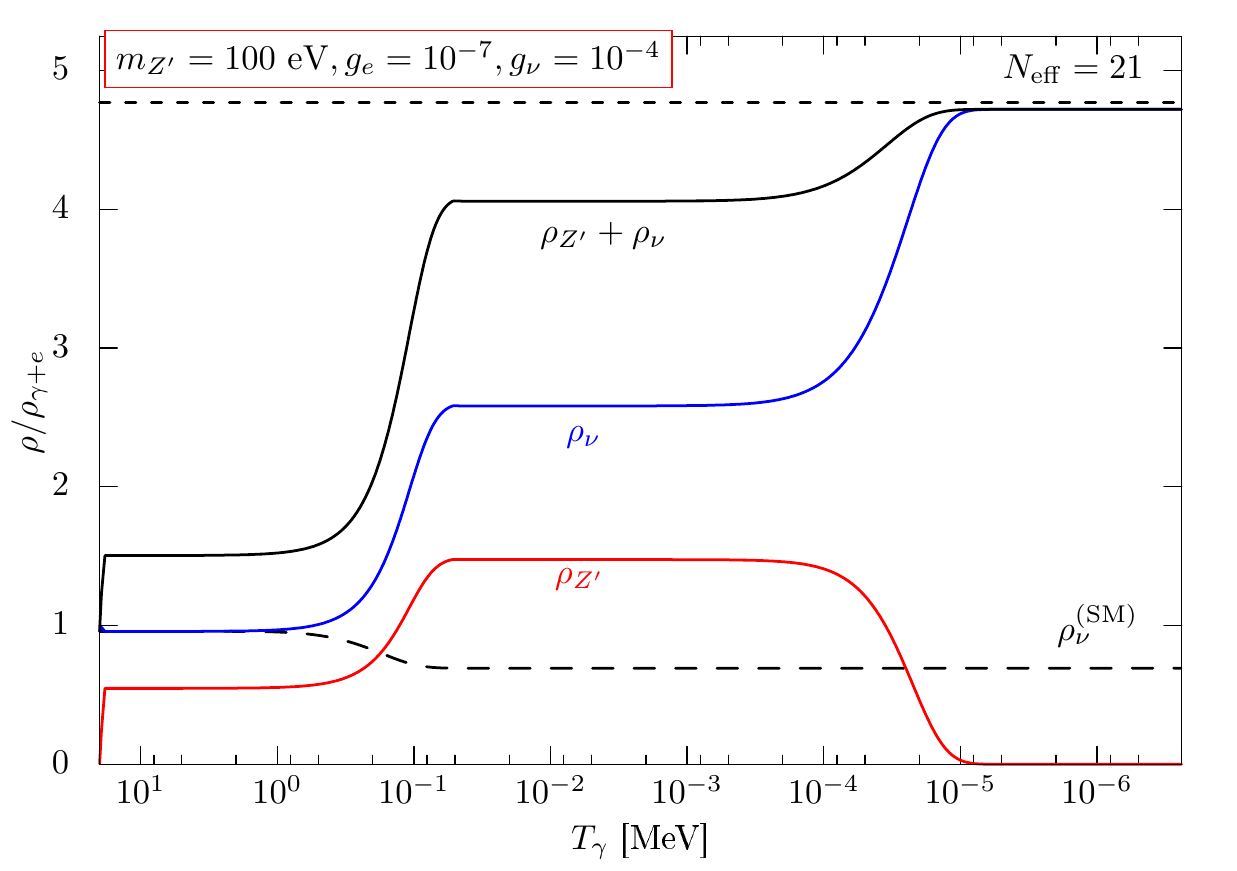}}
 	\subcaptionbox{Case (ii)\label{fig:case2} }
	{\includegraphics[width=0.47\textwidth]{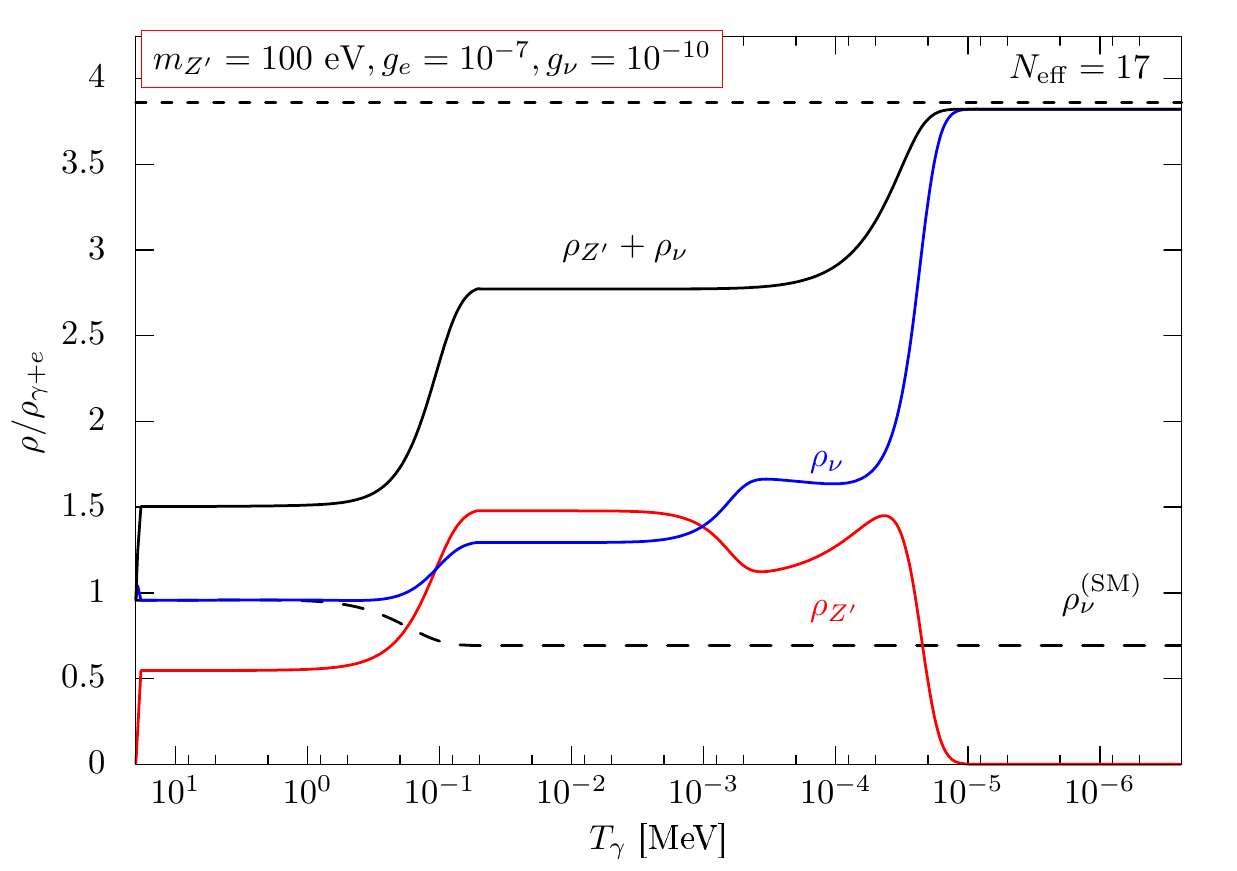}}
	
    \vspace{.4cm}

 	\subcaptionbox{Case (iii) \label{fig:case3} }
	{\includegraphics[width=0.47\textwidth]{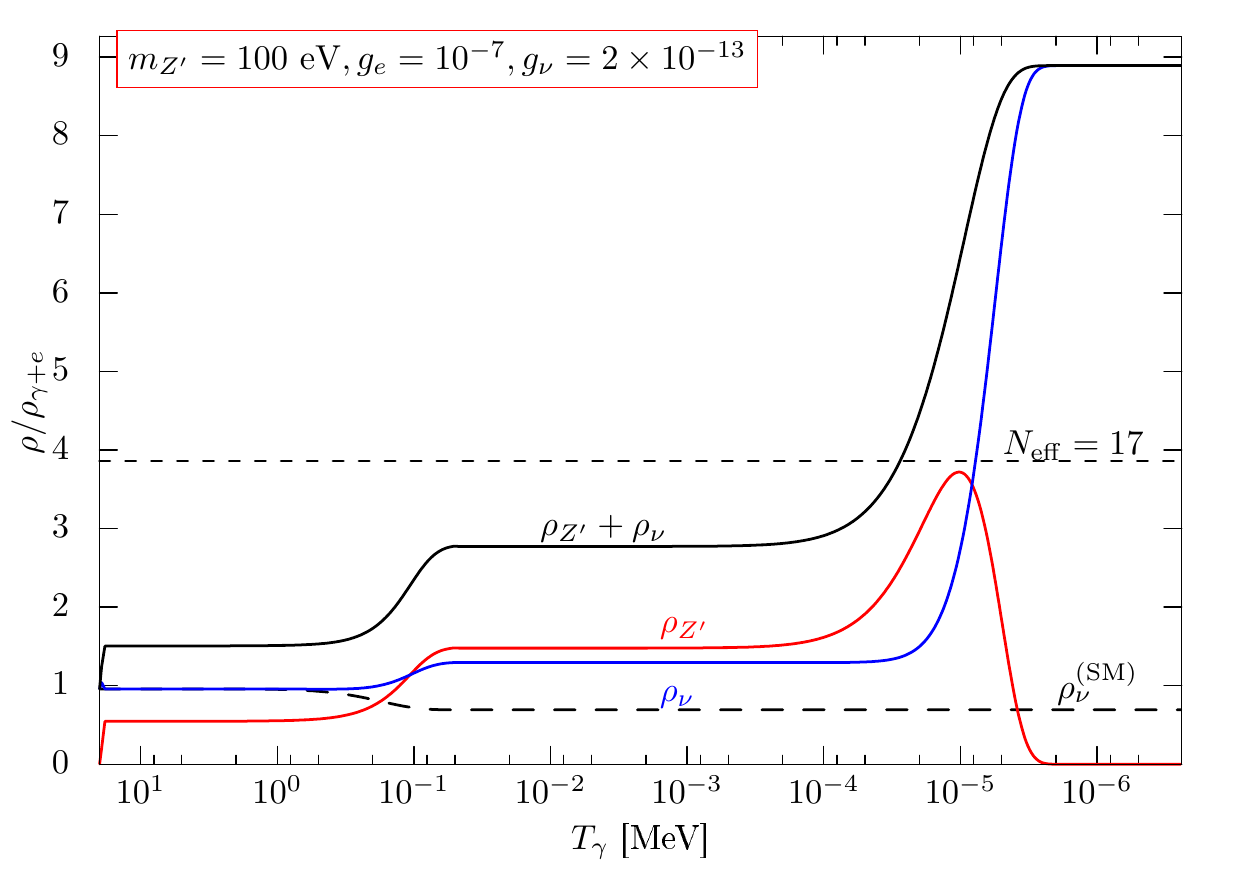}}
 	\subcaptionbox{Case (iv) \label{fig:case4}}
	{\includegraphics[width=0.47\textwidth]{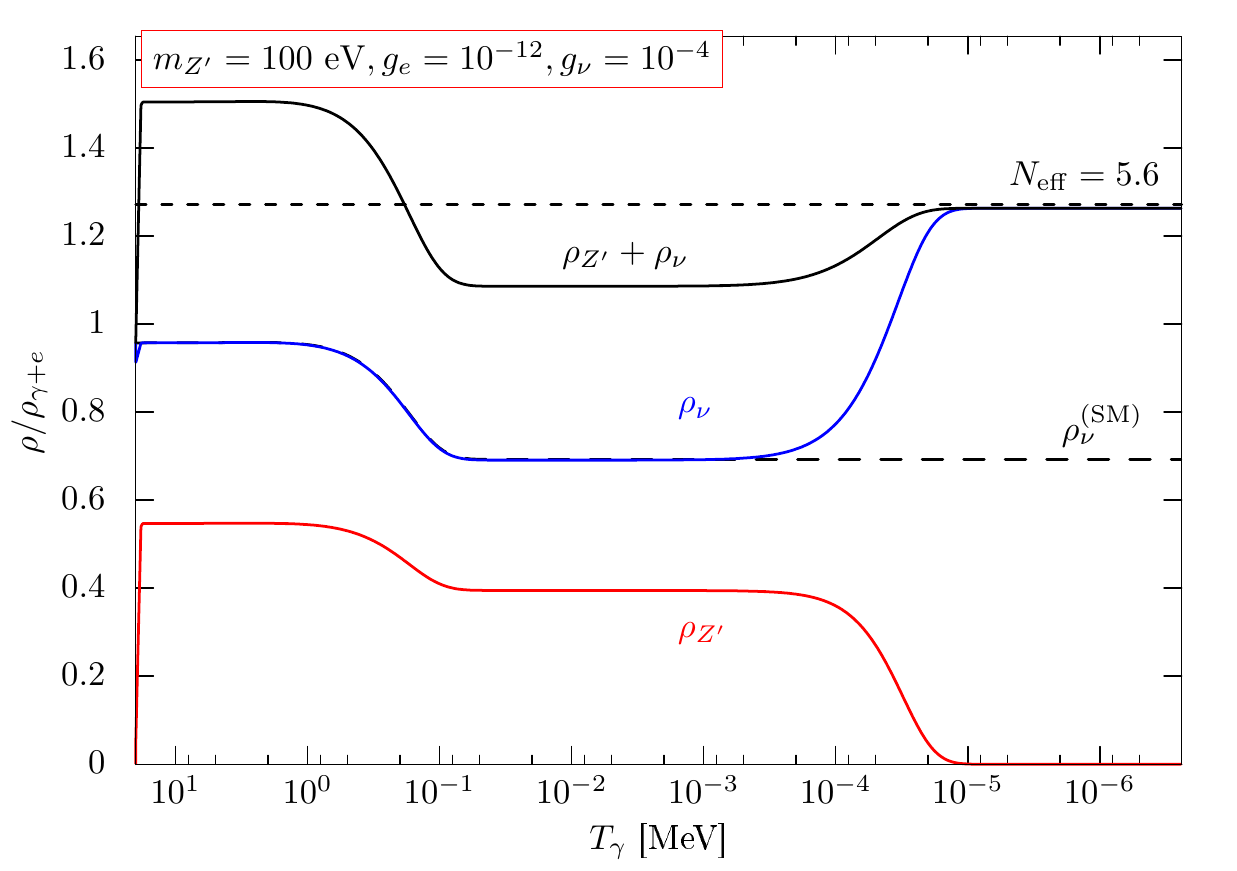}}
		
    \vspace{.4cm}
	
 		\subcaptionbox{Case (v)-1  \label{fig:case5-1}}
		{\includegraphics[width=0.47\textwidth]{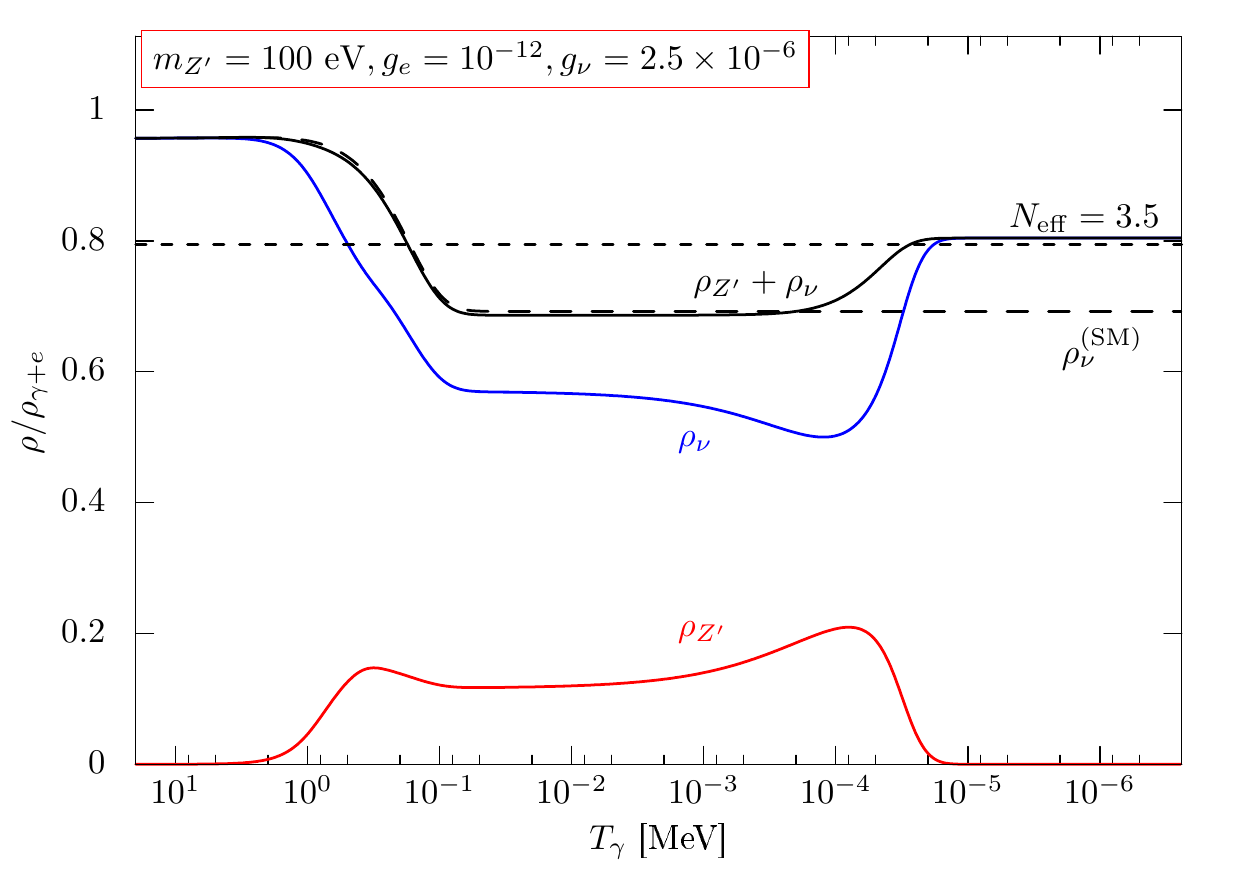}}
 	\subcaptionbox{Case (v)-2 \label{fig:case5-2}}
	{\includegraphics[width=0.47\textwidth]{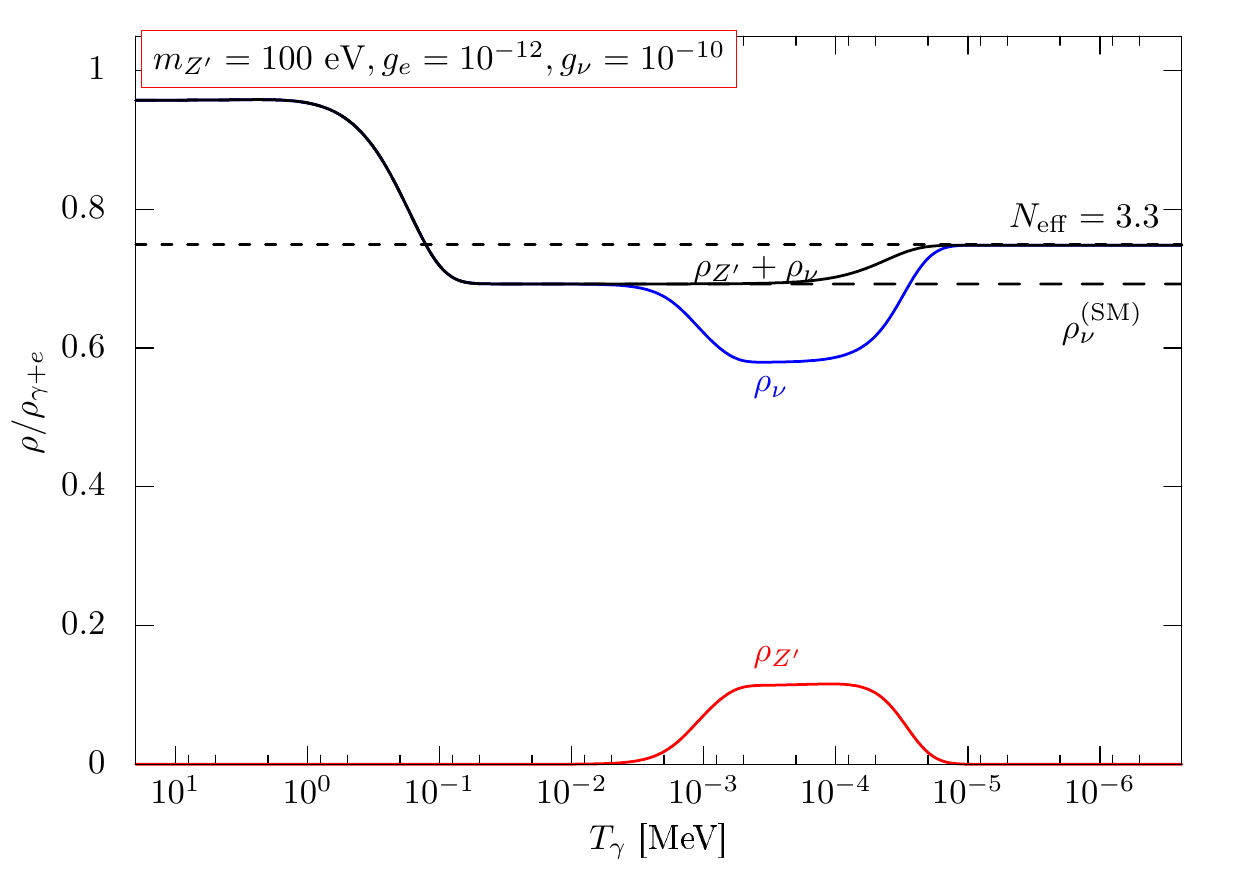}}
\caption{
Time evolution of the energy densities which show the similar behaviors of the Cases (i-v).
}\label{fig:cases}
\end{figure}

\newpage

\subsubsection*{(i) Simultaneous thermalization with both $e^\pm$ and $\nu$}
For $g_e \gg 10^{-8}$ and $g_\nu \gg 10^{-5}$, the on-shell productions of the mediator from the $\gamma\mbox{-} e$ thermal bath and the neutrino thermal bath are both effective, which delay the neutrino decoupling from the $\gamma \mbox{-} e$ thermal bath
until $T\lesssim m_e$.
As a result, $T_\nu \simeq T_\gamma$ is kept until 
$Z'$ annihilates away.
Then, the in-equilibrium decay of the mediator before the recombination heats up the neutrino 
temperature relative to the photon temperature by a factor of 
\begin{align}
\label{eq:neutrinoheating}
T_\nu = \left(\frac{\frac{7}{8}\times 2 N_F+d_{Z'}}{\frac{7}{8}\times 2 N_F}\right)^{1/3}T_\gamma = \left(\frac{11}{7}\right)^{1/3} T_\gamma\ ,
\end{align}
where $N_F = 3$ is the number of flavors of the neutrinos and $d_{Z'}=3$ is a spin degrees of freedom of $Z'$.
Here, we have used the conservation of the entropy per comoving volume in 
the $\nu\mbox{-}Z'$ thermal bath.
As a result, the expected $N_\mathrm{eff}$ for $g_e \gg 10^{-8}$ and $g_\nu \gg 10^{-5}$
is given by,
\begin{align}
    N_{\mathrm{eff}} \simeq 3 
    \times \left(\frac{11}{4}\right)^{4/3}
      \times \left(\frac{11}{7}\right)^{4/3}\simeq 21\ .
\end{align}
The first factor is due to the delay of the neutrino decoupling, i.e. $T_\nu \simeq T_\gamma$, while the second factor is due to the in-equilibrium decay of $Z'$.
In Fig.\,\ref{fig:case1}, we show the time evolution of the energy densities of $\nu$ and $Z'$
for $g_e = 10^{-7}$, $g_\nu = 10^{-4}$ and $m_{Z'} = 100$\,eV.

For $m_{Z'}\ll 1$\,eV, $Z'$ behaves as the dark radiation without heating up the $\nu$ temperature.
In such a case, the expected $N_{\mathrm{eff}}$ is given by,
\begin{align}
N_\mathrm{eff} \simeq  \left(\frac{11}{4}\right)^{4/3}\left(N_F+ \frac{1}{2}\times\frac{8}{7}\times d_{Z'}\right) \simeq 18\ ,
\end{align}
where the second term in the last parenthesis denotes the $Z'$ energy density.

\subsubsection*{(ii) Thermalization with $e^\pm$ followed by $\nu$-inverse decay}
For $g_e \gg 10^{-8}$ but for $g_\nu \ll 10^{-5}$ while satisfying the condition in Eq.\,\eqref{eq:inequilibrium}, the mediator remains in equilibrium with $\gamma\mbox{-}e$, even after the neutrino decouples as in the standard cosmology.
In this case, the temperature of the  $\gamma\mbox{-}Z'$ thermal bath after the electron annihilation is given by,
\begin{align}
T_\gamma = T_{Z'} = \left(\frac{\frac{7}{8}\times 4 + 2 +d_{Z'}}{2 +d_{Z'}}\right)^{1/3} T_\nu=\left(\frac{17}{10}\right)^{1/3} T_\nu\ .
\end{align}
After $e^\pm$ have annihilated away,
the (inverse) decay of $Z'$ into the neutrinos becomes effective.
The (inverse) decay changes the temperature and the chemical potential of $\nu\mbox{-}Z'$ thermal bath as  
\begin{align}
&\rho_\nu(T_\nu,0) + 
\rho_{Z'}((17/10)^{1/3}T_\nu,0)
= \rho_\nu(T_{\nu\mbox{-}Z'},\mu_{\nu\mbox{-}Z'}) + 
\rho_{Z'}(T_{\nu\mbox{-}Z'},2\mu_{\nu\mbox{-}Z'})\ , \\
\label{eq:nconserv}
&n_{\nu}(T_\nu,0)+ 
2n_{Z'}((17/10)^{1/3}T_\nu,0)
= n_\nu(T_{\nu\mbox{-}Z'},\mu_{\nu\mbox{-}Z'}) + 
2n_{Z'}(T_{\nu\mbox{-}Z'},2\mu_{\nu\mbox{-}Z'}) \ ,
\end{align}
for $\mu_{\nu\mbox{-}Z'} < 0$.
For $\mu_{\nu\mbox{-}Z'}\to 0$, on the other hand, the conditions are given by
\begin{align}
&\rho_\nu(T_\nu,0) + 
\rho_{Z'}((17/10)^{1/3}T_\nu,0)
= \rho_\nu(T_{\nu\mbox{-}Z'},0) + 
\rho_{Z'}(T_{\nu\mbox{-}Z'},0)\ , \\
\label{eq:nconservBEC}
&n_{\nu}(T_\nu,0)+ 
2n_{Z'}((17/10)^{1/3}T_\nu,0)
= n_\nu(T_{\nu\mbox{-}Z'},0) + 
2n_{Z'}(T_{\nu\mbox{-}Z'},0) + 
2n_{Z'}^{(0)}\ .
\end{align}
Here, $T_\nu$ denotes the neutrino temperature in the absence of the inverse decay of $Z'$. 
The mediator 
distribution is approximated by 
the Bose-Einstein distribution of the 
massless particle with the temperature $T$ and the chemical potential $\mu$.
The first and the second arguments of the energy/number densities are the temperature and the chemical potential, respectively.
The zero momentum contribution to the number density is denoted by $n_{Z'}^{(0)}$, of which the energy density is neglected in the massless approximation.
The conditions in Eqs.\,\eqref{eq:nconserv} and \eqref{eq:nconservBEC} are due to the conservation of $n_\nu + 2n_{Z'}$ in the (inverse) decay process, which also imposes $\mu_{Z'} = 2 \mu_{\nu}$.
The above conditions lead to
\begin{align}
\label{eq:caseii}
    T_{\nu\mbox{-}Z'} \simeq 1.1\, T_\nu\ ,
    \quad \mu_{\nu\mbox{-}Z'} = 0\ ,
    \quad n_{Z'}^{(0)} \simeq 0.084\,T_\nu^3\ .
\end{align}
The result in Eq.\,\eqref{eq:caseii} shows that $Z'$ exhibits a dilute Bose-Einstein condensation (BEC).
In the presence of the BEC, the zero momentum contribution should be treated properly in the Boltzmann equation (see, e.g., Refs.\,\cite{Semikoz:1994zp,Semikoz:1995rd}).
In our numerical analysis, however, we use the Boltzmann equation neglecting the zero momentum contribution.
Since $n_{Z'}^{(0)}$ is subdominant, our approximation fairly reproduces the expected $N_{\mathrm{eff}}$ discussed below (see also Fig.\,\ref{fig:case2}).

The in-equilibrium decay of $Z'$ 
before the CMB era heats up the neutrino temperature as
\begin{align}
\label{eq:sconserv}
&s_\nu(T_{\nu\mbox{-}Z'},0) + 
s_{Z'}(T_{\nu\mbox{-}Z'},0)
= s_\nu(T_{\nu}',\mu_{\nu}')\ ,\\
\label{eq:nconservDECAY}
&n_\nu(T_{\nu\mbox{-}Z'},0) + 
2n_{Z'}(T_{\nu\mbox{-}Z'},0) + 
2 n_{Z'}^{(0)}
= n_\nu(T_{\nu}',\mu_{\nu}')  \ .
\end{align}
The resultant temperature and the chemical potential of $\nu$ are given by,
\begin{align}
     T'_{\nu} \simeq 0.76\, T_\nu\ ,
    \quad \mu'_{\nu} \simeq 1.9 \,T_\nu \ ,
\end{align}
where $T_\nu$ is again the neutrino temperature in the absence of the inverse decay nor the decay of $Z'$.
As a result, we find%
\footnote{For $d_{Z'} = 1$, for example, we find $N_{\mathrm{eff}} = 12$ for $m_{Z'}\ge 1$\,eV 
and $N_{\mathrm{eff}} = 11$ for $m_{Z'}\ll 1$\,eV.}
\begin{align}
    N_{\mathrm{eff}} \simeq 17\ .
\end{align}
In Fig.\,\ref{fig:case2}, we show the time evolution of the energy densities of $\nu$ and $Z'$
for $g_e = 10^{-7}$, $g_\nu = 10^{-10}$ and $m_{Z'} = 100$\,eV.

For $m_{Z'} \ll 1$\,eV, 
the decay of $Z'$ takes place after the recombination, and hence, it contributes to $N_{\mathrm{eff}}$ as dark radiation as in the previous case.
The resultant $N_{\mathrm{eff} }$
is given by,
\begin{align}
    N_{\mathrm{eff}} \simeq 12 \ .
\end{align}

\subsubsection*{(iii) Thermalization with $e^\pm$ followed by out-of-equilibrium decay of $Z'$}
For $g_e \gg 10^{-9}$
and $\Gamma_{Z'} \ll H(T\simeq m_{Z'})$, 
 $Z'$ exhibits the non-equilibrium decay.
In this case, the $\nu$ energy density is more enhanced than that expected from Eqs.\,\eqref{eq:sconserv} and \eqref{eq:nconservDECAY}, and hence,
\begin{align}
    N_{\mathrm{eff}} > 17\ .
\end{align}
In Fig.\,\ref{fig:case3}, we show the time evolution of the energy densities of $\nu$ and $Z'$
for $g_e = 10^{-7}$, $g_\nu = 2\times 10^{-13}$ and $m_{Z'} = 100$\,eV.

For $m_{Z'}\ll 1$\,eV, $N_{\mathrm{eff}}$ is again given by,
\begin{align}
N_{\mathrm{eff}} \simeq 12\ ,
\end{align}
since $Z'$ is effectively an massless degree of freedom at the recombination.

\subsubsection*{(iv) Thermalization with $\nu$ before neutrino decoupling}
For $g_\nu \gg 10^{-5}$ and $g_e \ll 10^{-9}$, the production of $Z'$ from $e^{\pm}$ is not significant, while it is in equilibrium with the neutrino thermal bath.
When $Z'$ is thermalized with $\nu$
before the neutrino decoupling,
the temperature
of $\nu\mbox{-}Z'$ thermal bath is given by
\begin{align}
\label{eq:nuZtemp}
T_{\nu\mbox{-}Z'} \simeq \left(\frac{4}{11}\right)^{1/3}T_{\gamma} \ ,
\end{align}
with the vanishing chemical potential
after $e^\pm$ annihilates away.

After the in-equilibrium decay of $Z'$,
the neutrino temperature is enhanced by a factor in Eq.\,\eqref{eq:neutrinoheating},
\begin{align}
    T_\nu = \left(\frac{\frac{7}{8}\times 2 N_F+d_{Z'}}{\frac{7}{8}\times 2 N_F}\right)^{1/3}T_{\nu\mbox{-}Z'}=\left(\frac{11}{7}\right)^{1/3}T_{\nu\mbox{-}Z'}\ ,
\end{align}
where $T_{\nu\mbox{-}Z'}$ is the temperature without the in-equilibrium decay of $Z'$.
As a result, we obtain
\begin{align}
N_{\mathrm{eff}} \simeq N_{\mathrm{eff}}^{(
\mathrm{SM})} \times
\left(\frac{\frac{7}{8}\times 2 N_F+d_{Z'}}{\frac{7}{8}\times 2 N_F}\right)^{4/3}
\simeq 5.6\ . 
\end{align}
In Fig.\,\ref{fig:case4}, we show the time evolution of the energy densities of $\nu$ and $Z'$
for $g_e = 10^{-12}$, $g_\nu = 10^{-4}$ and $m_{Z'} = 100$\,eV.

For $m_{Z'}\ll 1$\,eV, the mediator contributes to $N_\mathrm{eff}$ as a dark radiation with the temperature in Eq.\,\eqref{eq:nuZtemp}.
As a result, $N_\mathrm{eff}$ is given by,
\begin{align}
    N_{\mathrm{eff}} \simeq N_{\mathrm{eff}}^{(
\mathrm{SM})} \times
\left(\frac{\frac{7}{8}\times 2 N_F+d_{Z'}}{\frac{7}{8}\times 2 N_F}\right)
\simeq 4.8\ .
\end{align}

\subsubsection*{(v) Thermalization with $\nu$ after neutrino decoupling}
For $g_e \ll 10^{-9}$ and $g_\nu \lesssim 10^{-5}$,
the thermalization with neutrino takes place after the neutrino decoupling.
Although such parameter region is far off from that favored by the XENON1T excess (see Fig.\,\ref{fig:Neff}), let us briefly comment on the behavior of $N_{\mathrm{eff}}$ in this region.

For $g_\nu$ with which both the $\nu+\nu \leftrightarrow Z'$ 
and $\nu+\nu \leftrightarrow Z'+Z'$ are effective (see Eqs.~\eqref{eq:nuscatter} and \eqref{eq:inequilibrium}), the chemical potentials of $\nu$ and $Z'$ vanish. 
Hence, the energy density of $\nu$ and $Z'$ after thermalization is determined by,
\begin{align}
    \rho_{\nu}(T_\nu,0) =
    \rho_{\nu}(T_{\nu\mbox{-}Z'},0) + \rho_{Z'}(T_{\nu\mbox{-}Z'},0)\ .
\end{align}
Thus, the temperature of the $\nu$-$Z'$ thermal bath is given by,
\begin{align}
T_{\nu\mbox{-}Z'} = \left(\frac{\frac{7}{8}\times 2 N_F}{\frac{7}{8}\times 2 N_F+d_{Z'}}\right)^{1/4}T_\nu=\left(\frac{7}{11}\right)^{1/4}T_\nu\ .
\end{align}
After the in-equilibrium decay of $Z'$, the neutrino temperature is heat up by a factor of $(11/7)^{1/3}$.
As a result, we find 
\begin{align}
    N_{\mathrm{eff}} \simeq N_{\mathrm{eff}}^{(\mathrm{SM})}\times \left(\frac{7}{11}\right)\times \left(\frac{11}{7}\right)^{4/3} \simeq 3.5\ .
\end{align}
In Fig.\,\ref{fig:case5-1}, we show the time evolution of the energy densities of $\nu$ and $Z'$
for $g_e = 10^{-12}$, $g_\nu = 2.5\times 10^{-6}$ and $m_{Z'} = 100$\,eV.

For $g_\nu \ll 10^{-5}$ but satisfying the condition in Eq.\,\eqref{eq:inequilibrium}, $\nu+\nu \leftrightarrow Z'+Z'$ is irrelevant, while $\nu+\nu \leftrightarrow Z'$ is effective.
In this case, the temperature and the chemical potential of the $\nu\mbox{-}Z'$ thermal bath are determined by
\begin{align}
\label{eq:rhoconserv2}
&\rho_\nu(T_\nu,0) 
= \rho_\nu(T_{\nu\mbox{-}Z'},\mu_{\nu\mbox{-}Z'}) + 
\rho_{Z'}(T_{\nu\mbox{-}Z'},2\mu_{\nu\mbox{-}Z'})\ ,\\
\label{eq:nconserv2}
&n_{\nu}(T_\nu,0)
= n_\nu(T_{\nu\mbox{-}Z'},\mu_{\nu\mbox{-}Z'}) + 
2n_{Z'}(T_{\nu\mbox{-}Z'},2\mu_{\nu\mbox{-}Z'}) \ .
\end{align}
The resultant temperature and the chemical potential of $\nu$ are given by,
\begin{align}
      T_{\nu\mbox{-}Z'} \simeq 1.2\, T_\nu\ ,
    \quad \mu_{\nu\mbox{-}Z'} \simeq -1.2 \,T_\nu \ .
\end{align}
Here, $T_\nu$ is again the neutrino temperature in the absence of the  (inverse) decay of $Z'$.

Then, the in-equilibrium decay of $Z'$ 
before the recombination heats up the neutrino temperature as,
\begin{align}
\label{eq:sconserv3}
&s_\nu(T_{\nu\mbox{-}Z'},\mu_{\nu\mbox{-}Z'}) + 
s_{Z'}(T_{\nu\mbox{-}Z'},2\mu_{\nu\mbox{-}Z'})
= s_\nu(T_{\nu}',\mu_{\nu}')\ ,\\
\label{eq:nconserv3}
&n_\nu(T_{\nu\mbox{-}Z'},\mu_{\nu\mbox{-}Z'}) + 
2n_{Z'}(T_{\nu\mbox{-}Z'},2\mu_{\nu\mbox{-}Z'})
= n_\nu(T_{\nu}',\mu_{\nu}')  \ .
\end{align}
The resultant temperature and the chemical potential of $\nu$ are given by,
\begin{align}
           T'_{\nu} \simeq 1.1\, T_\nu\ ,
    \quad \mu'_{\nu} \simeq -0.31 \,T_\nu \  .
\end{align}
As a result, we find
\begin{align}
    N_{\mathrm{eff}} \simeq 3.3\ ,
\end{align}
which is consistent with the joint Planck+BAO constraint on $N_{\mathrm{eff}}$ at 95\% C.L.,
while it is excluded by the CMB only constraint at 95\% C.L.
In Fig.\,\ref{fig:case5-2}, we show the time evolution of the energy densities of $\nu$ and $Z'$
for $g_e = 10^{-12}$, $g_\nu = 10^{-10}$ and $m_{Z'} = 100$\,eV.

\begin{figure}[t]
\centering
	\includegraphics[width=0.8\hsize,clip]{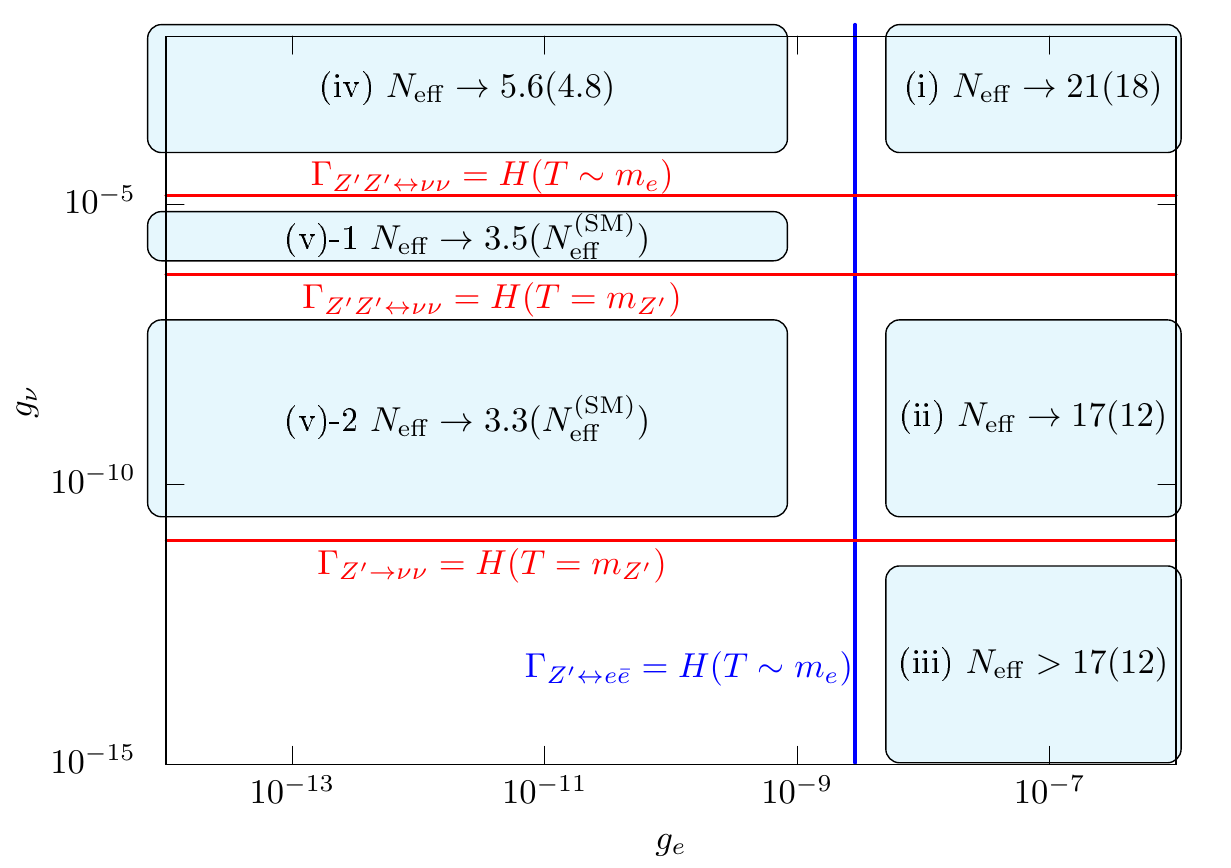}

	\caption{
	Schematic picture of summary of the each case study of the section~\ref{sec:Neffconstraint}.}
    \label{fig:summary}
\end{figure}

For $m_{Z'} \ll 1$\,eV, the mediator $Z'$ contributes to $N_\mathrm{eff}$ as a dark radiation.
However, due to the energy conservation in Eq.\,\eqref{eq:rhoconserv2}, the total energy density of the $\nu+Z'$ is the same with that of $\nu$ in the standard cosmology, and hence,
\begin{align}
    N_{\mathrm{eff}} = N_{\mathrm{eff}}^{(\mathrm{SM})}\ .
\end{align}
Thus, for $m_{Z'}\ll 1$\,eV, $g_\nu\ll 10^{-5}$ and $g_{e}\ll 10^{-9}$, the mediator is not constrained by the $N_{\mathrm{eff}}$ observation, even if the mediator is thermalized by the inverse decay of the neutrinos. 
We summarize the each case of the above and the corresponding asymptotic values of $N_\mathrm{eff}$ in Fig.~\ref{fig:summary}.

It should be noted that the cases (i)--(iv) are also constrained by the Big-Bang Nucleosynthesis (BBN) since the energy density  $\rho_\nu+\rho_{Z'}$ deviates from $\rho_\nu$ in the standard cosmology at $T_\gamma < \order{0.1}$\,MeV.
Typically, the BBN constraint 
on $|{\mit \Delta}N_{\mathrm{eff}}|$
is about $0.3$ 
at 68\%\,C.L.~\cite{Pitrou:2018cgg,Fields:2019pfx}, which is slightly weaker than the CMB constraint.
In the case (v), on the other hand,
the energy density  $\rho_\nu+\rho_{Z'}$ is the same with 
$\rho_\nu$ in the standard cosmology
for $T_\nu\gg m_{Z'}$ (see e.g. Ref.\,\cite{Berlin:2019pbq}).
Thus, the BBN constraints can be evaded for a very light $Z'$.
As we have seen, for such a very light $Z'$ with $m_{Z'}\lesssim {100}$\,eV, there is a constraint from the neutrino free-streaming on the CMB  in which the interaction between $Z'$ and the neutrinos reduces the free-streaming length~\cite{Chacko:2003dt,Escudero:2019gvw}. 

\section{Conclusions}
\label{sec:Discussion}
In this paper, we discussed
the constraints on the light vector mediator interpretation of the XENON1T excess from 
$N_\mathrm{eff}$.
This interpretation favors the mediator 
with a mass $m_{Z'}< \order{100}$\,keV 
and the electron/neutrino coupling $(g_eg_\nu)^{1/2}\simeq 3 \times 10^{-7}$~\cite{Boehm:2020ltd}.
By solving the Boltzmann equation of the momentum distribution of the mediator, we find that the favored parameter region results in $N_{\mathrm{eff}} > 5$ for $m_{Z'} \ge 1$\,eV, which exceeds the current upper limit on $N_{\mathrm{eff}}$.
By combined with the very conservative constraint from the stellar cooling on the light mediator coupling to the electron in Ref.\,\cite{Redondo:2008aa,An_2013},
we conclude that the light vector mediator interpretation is not valid anymore.

In this paper, we considered the case of the vector mediators.
To explain the XENON1T excess, the scalar mediator models have been also proposed.
For those model, we can apply $N_{\mathrm{eff} }$ constraint in a similar manner.
However, the thermal history of those cases depends on the detail of the models such as the presence of the right-handed neutrinos.
Although detailed analyses are out of scope of this work, 
our results of 
the thermalized cases (i)-(iv) can be applied straightforwardly 
by replacing $d_{Z'} = 3$ to $d_{Z'}=1$.
In these cases, the mediator couplings favored by the XENON1T excess seem to be in strong tension with the $N_{\mathrm{eff}}$ constraint.
For the detail analysis, we will discuss these cases elsewhere.

\section*{Acknowledgments}

This work is supported by Grant-in-Aid for Scientific Research from the Ministry of Education, Culture, Sports, Science, and Technology (MEXT), Japan, 17H02878 (M.I. and S.S.), 18H05542 (M.I.), 18K13535, 19H04609 and 20H01895 (S.S.), and by World Premier International Research Center Initiative (WPI), MEXT, Japan. 
This work is also supported by the Advanced Leading Graduate Course for Photon Science (S.K.), the JSPS Research Fellowships for Young Scientists (S.K.) and International Graduate Program for Excellence in Earth-Space Science (Y.N.).

\appendix
\section{Neutrino-Electron Collision Terms}
\label{sec:neutrino collision}
Here, we present the explicit form of the collision terms for neutrino-electron scatterings used in this work.
In the text, we define the collision terms of these processes for the zeroth and first moment
\begin{align}
    C^{(j)}_{e \leftrightarrow \nu} &= \sum_{i = e, \mu, \tau} C^{(j)}_{e \leftrightarrow \nu_i},  \, \, \, \, j = 0, 1\\
    C^{(0)}_{e \leftrightarrow \nu_i} &= \int \frac{g_{\nu_i} d^3 p_{\nu_i}}{(2\pi)^3}
    \qty(\mathcal{C}_{e^+e^-\leftrightarrow \nu_i \bar{\nu}_i} +\mathcal{C}_{e^\pm\nu_i\leftrightarrow e^\pm\nu_i}
    +\mathcal{C}_{e^\pm\bar{\nu}_i\leftrightarrow e^\pm\bar{\nu}_i})\ , \\
    C^{(1)}_{e \leftrightarrow \nu_i} &= \int \frac{g_{\nu_i} d^3 p_{\nu_i}}{(2\pi)^3} p_{\nu_i}
    \qty(\mathcal{C}_{e^+e^-\leftrightarrow \nu_i \bar{\nu}_i} +\mathcal{C}_{e^\pm\nu_i\leftrightarrow e^\pm\nu_i} +\mathcal{C}_{e^\pm\bar{\nu}_i\leftrightarrow e^\pm\bar{\nu}_i})\ .
\end{align}
Using the results of Appendix A.3 of Ref. \cite{Escudero:2020dfa}, these are written in the form of
\begin{align}
    C^{(0)}_{e\leftrightarrow\nu} &= 8f_n^{\rm FD}\frac{G_F^2}{\pi^5}(3-4s_{\rm W}^2 + 24s_{\rm W}^4)(T_{\gamma}^8 - T_{\nu}^8
    e^{\frac{2\mu_{\nu}}{T_{\nu}}})\ , \\
    C^{(1)}_{e\leftrightarrow\nu} &= \frac{G_F^2}{\pi^5}(3-4s_{\rm W}^2 + 24s_{\rm W}^4) G(T_{\gamma}, 0, T_{\nu}, \mu_{\nu})\ ,
\end{align}
where $G_F$ is the Fermi constant, $s_W$ is the sine of the Weinberg angle, and $f_n^{\rm FD} = 0.852$. 
The function $G(T_1, \mu_1, T_2, \mu_2)$ is given by
\begin{align}
    G(T_1, \mu_1, T_2, \mu_2) = 32f_a^{\rm FD}(T_1^9e^{2\frac{\mu_1}{T_1}} - T_2^9e^{2\frac{\mu_2}{T_2}})
    + 56f_s^{\rm FD}e^{\frac{\mu_1}{T_1}}e^{\frac{\mu_2}{T_2}}T_1^4 T_2^4 (T_1 - T_2)\ ,
\end{align}
where the numerical factors $f_a^{\rm FD} = 0.884$ and $f_s^{\rm FD} = 0.829$ represent the Pauli blocking effect from the Fermi-Dirac distribution in the annihilation and scattering processes, respectively. Note that we assume that each neutrino flavor has the same temperature and chemical potential. 

\bibliographystyle{apsrev4-1}
\bibliography{ref}

\begin{thebibliography}{37}%
\makeatletter
\providecommand \@ifxundefined [1]{%
 \@ifx{#1\undefined}
}%
\providecommand \@ifnum [1]{%
 \ifnum #1\expandafter \@firstoftwo
 \else \expandafter \@secondoftwo
 \fi
}%
\providecommand \@ifx [1]{%
 \ifx #1\expandafter \@firstoftwo
 \else \expandafter \@secondoftwo
 \fi
}%
\providecommand \natexlab [1]{#1}%
\providecommand \enquote  [1]{``#1''}%
\providecommand \bibnamefont  [1]{#1}%
\providecommand \bibfnamefont [1]{#1}%
\providecommand \citenamefont [1]{#1}%
\providecommand \href@noop [0]{\@secondoftwo}%
\providecommand \href [0]{\begingroup \@sanitize@url \@href}%
\providecommand \@href[1]{\@@startlink{#1}\@@href}%
\providecommand \@@href[1]{\endgroup#1\@@endlink}%
\providecommand \@sanitize@url [0]{\catcode `\\12\catcode `\$12\catcode
  `\&12\catcode `\#12\catcode `\^12\catcode `\_12\catcode `\%12\relax}%
\providecommand \@@startlink[1]{}%
\providecommand \@@endlink[0]{}%
\providecommand \url  [0]{\begingroup\@sanitize@url \@url }%
\providecommand \@url [1]{\endgroup\@href {#1}{\urlprefix }}%
\providecommand \urlprefix  [0]{URL }%
\providecommand \Eprint [0]{\href }%
\providecommand \doibase [0]{http://dx.doi.org/}%
\providecommand \selectlanguage [0]{\@gobble}%
\providecommand \bibinfo  [0]{\@secondoftwo}%
\providecommand \bibfield  [0]{\@secondoftwo}%
\providecommand \translation [1]{[#1]}%
\providecommand \BibitemOpen [0]{}%
\providecommand \bibitemStop [0]{}%
\providecommand \bibitemNoStop [0]{.\EOS\space}%
\providecommand \EOS [0]{\spacefactor3000\relax}%
\providecommand \BibitemShut  [1]{\csname bibitem#1\endcsname}%
\let\auto@bib@innerbib\@empty
\bibitem [{\citenamefont {Aprile}\ \emph {et~al.}(2020)\citenamefont {Aprile}
  \emph {et~al.}}]{Aprile:2020tmw}%
  \BibitemOpen
  \bibfield  {author} {\bibinfo {author} {\bibfnamefont {E.}~\bibnamefont
  {Aprile}} \emph {et~al.} (\bibinfo {collaboration} {XENON}),\ }\href@noop {}
  {\  (\bibinfo {year} {2020})},\ \Eprint {http://arxiv.org/abs/2006.09721}
  {arXiv:2006.09721 [hep-ex]} \BibitemShut {NoStop}%
\bibitem [{\citenamefont {Di~Luzio}\ \emph {et~al.}(2020)\citenamefont
  {Di~Luzio}, \citenamefont {Fedele}, \citenamefont {Giannotti}, \citenamefont
  {Mescia},\ and\ \citenamefont {Nardi}}]{DiLuzio:2020jjp}%
  \BibitemOpen
  \bibfield  {author} {\bibinfo {author} {\bibfnamefont {L.}~\bibnamefont
  {Di~Luzio}}, \bibinfo {author} {\bibfnamefont {M.}~\bibnamefont {Fedele}},
  \bibinfo {author} {\bibfnamefont {M.}~\bibnamefont {Giannotti}}, \bibinfo
  {author} {\bibfnamefont {F.}~\bibnamefont {Mescia}}, \ and\ \bibinfo {author}
  {\bibfnamefont {E.}~\bibnamefont {Nardi}},\ }\href@noop {} {\  (\bibinfo
  {year} {2020})},\ \Eprint {http://arxiv.org/abs/2006.12487} {arXiv:2006.12487
  [hep-ph]} \BibitemShut {NoStop}%
\bibitem [{\citenamefont {Boehm}\ \emph {et~al.}(2020)\citenamefont {Boehm},
  \citenamefont {Cerdeno}, \citenamefont {Fairbairn}, \citenamefont {Machado},\
  and\ \citenamefont {Vincent}}]{Boehm:2020ltd}%
  \BibitemOpen
  \bibfield  {author} {\bibinfo {author} {\bibfnamefont {C.}~\bibnamefont
  {Boehm}}, \bibinfo {author} {\bibfnamefont {D.~G.}\ \bibnamefont {Cerdeno}},
  \bibinfo {author} {\bibfnamefont {M.}~\bibnamefont {Fairbairn}}, \bibinfo
  {author} {\bibfnamefont {P.~A.}\ \bibnamefont {Machado}}, \ and\ \bibinfo
  {author} {\bibfnamefont {A.~C.}\ \bibnamefont {Vincent}},\ }\href@noop {} {\
  (\bibinfo {year} {2020})},\ \Eprint {http://arxiv.org/abs/2006.11250}
  {arXiv:2006.11250 [hep-ph]} \BibitemShut {NoStop}%
\bibitem [{\citenamefont {Amaral}\ \emph {et~al.}(2020)\citenamefont {Amaral},
  \citenamefont {Cerdeno}, \citenamefont {Foldenauer},\ and\ \citenamefont
  {Reid}}]{Amaral:2020tga}%
  \BibitemOpen
  \bibfield  {author} {\bibinfo {author} {\bibfnamefont {d.}~\bibnamefont
  {Amaral}, \bibfnamefont {Dorian Warren~Praia}}, \bibinfo {author}
  {\bibfnamefont {D.~G.}\ \bibnamefont {Cerdeno}}, \bibinfo {author}
  {\bibfnamefont {P.}~\bibnamefont {Foldenauer}}, \ and\ \bibinfo {author}
  {\bibfnamefont {E.}~\bibnamefont {Reid}},\ }\href@noop {} {\  (\bibinfo
  {year} {2020})},\ \Eprint {http://arxiv.org/abs/2006.11225} {arXiv:2006.11225
  [hep-ph]} \BibitemShut {NoStop}%
\bibitem [{\citenamefont {Bally}\ \emph {et~al.}(2020)\citenamefont {Bally},
  \citenamefont {Jana},\ and\ \citenamefont {Trautner}}]{Bally:2020yid}%
  \BibitemOpen
  \bibfield  {author} {\bibinfo {author} {\bibfnamefont {A.}~\bibnamefont
  {Bally}}, \bibinfo {author} {\bibfnamefont {S.}~\bibnamefont {Jana}}, \ and\
  \bibinfo {author} {\bibfnamefont {A.}~\bibnamefont {Trautner}},\ }\href@noop
  {} {\  (\bibinfo {year} {2020})},\ \Eprint {http://arxiv.org/abs/2006.11919}
  {arXiv:2006.11919 [hep-ph]} \BibitemShut {NoStop}%
\bibitem [{\citenamefont {Aristizabal~Sierra}\ \emph
  {et~al.}(2020)\citenamefont {Aristizabal~Sierra}, \citenamefont {De~Romeri},
  \citenamefont {Flores},\ and\ \citenamefont
  {Papoulias}}]{AristizabalSierra:2020edu}%
  \BibitemOpen
  \bibfield  {author} {\bibinfo {author} {\bibfnamefont {D.}~\bibnamefont
  {Aristizabal~Sierra}}, \bibinfo {author} {\bibfnamefont {V.}~\bibnamefont
  {De~Romeri}}, \bibinfo {author} {\bibfnamefont {L.}~\bibnamefont {Flores}}, \
  and\ \bibinfo {author} {\bibfnamefont {D.}~\bibnamefont {Papoulias}},\
  }\href@noop {} {\  (\bibinfo {year} {2020})},\ \Eprint
  {http://arxiv.org/abs/2006.12457} {arXiv:2006.12457 [hep-ph]} \BibitemShut
  {NoStop}%
\bibitem [{\citenamefont {Khan}(2020)}]{Khan:2020vaf}%
  \BibitemOpen
  \bibfield  {author} {\bibinfo {author} {\bibfnamefont {A.~N.}\ \bibnamefont
  {Khan}},\ }\href@noop {} {\  (\bibinfo {year} {2020})},\ \Eprint
  {http://arxiv.org/abs/2006.12887} {arXiv:2006.12887 [hep-ph]} \BibitemShut
  {NoStop}%
\bibitem [{\citenamefont {Lindner}\ \emph {et~al.}(2020)\citenamefont
  {Lindner}, \citenamefont {Mambrini}, \citenamefont {Melo},\ and\
  \citenamefont {Queiroz}}]{Lindner:2020kko}%
  \BibitemOpen
  \bibfield  {author} {\bibinfo {author} {\bibfnamefont {M.}~\bibnamefont
  {Lindner}}, \bibinfo {author} {\bibfnamefont {Y.}~\bibnamefont {Mambrini}},
  \bibinfo {author} {\bibfnamefont {T.~B.~d.}\ \bibnamefont {Melo}}, \ and\
  \bibinfo {author} {\bibfnamefont {F.~S.}\ \bibnamefont {Queiroz}},\
  }\href@noop {} {\  (\bibinfo {year} {2020})},\ \Eprint
  {http://arxiv.org/abs/2006.14590} {arXiv:2006.14590 [hep-ph]} \BibitemShut
  {NoStop}%
\bibitem [{\citenamefont {Díaz}\ \emph {et~al.}(2019)\citenamefont {Díaz},
  \citenamefont {Schröder}, \citenamefont {Zuber}, \citenamefont {Jack},\ and\
  \citenamefont {Barrios}}]{daz2019constraint}%
  \BibitemOpen
  \bibfield  {author} {\bibinfo {author} {\bibfnamefont {S.~A.}\ \bibnamefont
  {Díaz}}, \bibinfo {author} {\bibfnamefont {K.-P.}\ \bibnamefont
  {Schröder}}, \bibinfo {author} {\bibfnamefont {K.}~\bibnamefont {Zuber}},
  \bibinfo {author} {\bibfnamefont {D.}~\bibnamefont {Jack}}, \ and\ \bibinfo
  {author} {\bibfnamefont {E.~E.~B.}\ \bibnamefont {Barrios}},\ }\href@noop {}
  {\enquote {\bibinfo {title} {Constraint on the axion-electron coupling
  constant and the neutrino magnetic dipole moment by using the tip-rgb
  luminosity of fifty globular clusters},}\ } (\bibinfo {year} {2019}),\
  \Eprint {http://arxiv.org/abs/1910.10568} {arXiv:1910.10568 [astro-ph.SR]}
  \BibitemShut {NoStop}%
\bibitem [{\citenamefont {Grifols}\ and\ \citenamefont
  {Masso}(1986)}]{Grifols:1986fc}%
  \BibitemOpen
  \bibfield  {author} {\bibinfo {author} {\bibfnamefont {J.}~\bibnamefont
  {Grifols}}\ and\ \bibinfo {author} {\bibfnamefont {E.}~\bibnamefont
  {Masso}},\ }\href {\doibase 10.1016/0370-2693(86)90509-5} {\bibfield
  {journal} {\bibinfo  {journal} {Phys. Lett. B}\ }\textbf {\bibinfo {volume}
  {173}},\ \bibinfo {pages} {237} (\bibinfo {year} {1986})}\BibitemShut
  {NoStop}%
\bibitem [{\citenamefont {Grifols}\ \emph {et~al.}(1989)\citenamefont
  {Grifols}, \citenamefont {Masso},\ and\ \citenamefont
  {Peris}}]{Grifols:1988fv}%
  \BibitemOpen
  \bibfield  {author} {\bibinfo {author} {\bibfnamefont {J.}~\bibnamefont
  {Grifols}}, \bibinfo {author} {\bibfnamefont {E.}~\bibnamefont {Masso}}, \
  and\ \bibinfo {author} {\bibfnamefont {S.}~\bibnamefont {Peris}},\ }\href
  {\doibase 10.1142/S0217732389000381} {\bibfield  {journal} {\bibinfo
  {journal} {Mod. Phys. Lett. A}\ }\textbf {\bibinfo {volume} {4}},\ \bibinfo
  {pages} {311} (\bibinfo {year} {1989})}\BibitemShut {NoStop}%
\bibitem [{\citenamefont {An}\ \emph {et~al.}(2013)\citenamefont {An},
  \citenamefont {Pospelov},\ and\ \citenamefont {Pradler}}]{An_2013}%
  \BibitemOpen
  \bibfield  {author} {\bibinfo {author} {\bibfnamefont {H.}~\bibnamefont
  {An}}, \bibinfo {author} {\bibfnamefont {M.}~\bibnamefont {Pospelov}}, \ and\
  \bibinfo {author} {\bibfnamefont {J.}~\bibnamefont {Pradler}},\ }\href
  {\doibase 10.1016/j.physletb.2013.07.008} {\bibfield  {journal} {\bibinfo
  {journal} {Physics Letters B}\ }\textbf {\bibinfo {volume} {725}},\ \bibinfo
  {pages} {190–195} (\bibinfo {year} {2013})}\BibitemShut {NoStop}%
\bibitem [{\citenamefont {Redondo}\ and\ \citenamefont
  {Raffelt}(2013)}]{Redondo_2013}%
  \BibitemOpen
  \bibfield  {author} {\bibinfo {author} {\bibfnamefont {J.}~\bibnamefont
  {Redondo}}\ and\ \bibinfo {author} {\bibfnamefont {G.}~\bibnamefont
  {Raffelt}},\ }\href {\doibase 10.1088/1475-7516/2013/08/034} {\bibfield
  {journal} {\bibinfo  {journal} {Journal of Cosmology and Astroparticle
  Physics}\ }\textbf {\bibinfo {volume} {2013}},\ \bibinfo {pages} {034–034}
  (\bibinfo {year} {2013})}\BibitemShut {NoStop}%
\bibitem [{\citenamefont {Chang}\ \emph {et~al.}(2017)\citenamefont {Chang},
  \citenamefont {Essig},\ and\ \citenamefont {McDermott}}]{Chang:2016ntp}%
  \BibitemOpen
  \bibfield  {author} {\bibinfo {author} {\bibfnamefont {J.~H.}\ \bibnamefont
  {Chang}}, \bibinfo {author} {\bibfnamefont {R.}~\bibnamefont {Essig}}, \ and\
  \bibinfo {author} {\bibfnamefont {S.~D.}\ \bibnamefont {McDermott}},\ }\href
  {\doibase 10.1007/JHEP01(2017)107} {\bibfield  {journal} {\bibinfo  {journal}
  {JHEP}\ }\textbf {\bibinfo {volume} {01}},\ \bibinfo {pages} {107} (\bibinfo
  {year} {2017})},\ \Eprint {http://arxiv.org/abs/1611.03864} {arXiv:1611.03864
  [hep-ph]} \BibitemShut {NoStop}%
\bibitem [{\citenamefont {Chang}\ \emph {et~al.}(2018)\citenamefont {Chang},
  \citenamefont {Essig},\ and\ \citenamefont {McDermott}}]{Chang:2018rso}%
  \BibitemOpen
  \bibfield  {author} {\bibinfo {author} {\bibfnamefont {J.~H.}\ \bibnamefont
  {Chang}}, \bibinfo {author} {\bibfnamefont {R.}~\bibnamefont {Essig}}, \ and\
  \bibinfo {author} {\bibfnamefont {S.~D.}\ \bibnamefont {McDermott}},\ }\href
  {\doibase 10.1007/JHEP09(2018)051} {\bibfield  {journal} {\bibinfo  {journal}
  {JHEP}\ }\textbf {\bibinfo {volume} {09}},\ \bibinfo {pages} {051} (\bibinfo
  {year} {2018})},\ \Eprint {http://arxiv.org/abs/1803.00993} {arXiv:1803.00993
  [hep-ph]} \BibitemShut {NoStop}%
\bibitem [{\citenamefont {Redondo}(2008)}]{Redondo:2008aa}%
  \BibitemOpen
  \bibfield  {author} {\bibinfo {author} {\bibfnamefont {J.}~\bibnamefont
  {Redondo}},\ }\href {\doibase 10.1088/1475-7516/2008/07/008} {\bibfield
  {journal} {\bibinfo  {journal} {JCAP}\ }\textbf {\bibinfo {volume} {07}},\
  \bibinfo {pages} {008} (\bibinfo {year} {2008})},\ \Eprint
  {http://arxiv.org/abs/0801.1527} {arXiv:0801.1527 [hep-ph]} \BibitemShut
  {NoStop}%
\bibitem [{\citenamefont {de~Salas}\ and\ \citenamefont
  {Pastor}(2016)}]{deSalas:2016ztq}%
  \BibitemOpen
  \bibfield  {author} {\bibinfo {author} {\bibfnamefont {P.~F.}\ \bibnamefont
  {de~Salas}}\ and\ \bibinfo {author} {\bibfnamefont {S.}~\bibnamefont
  {Pastor}},\ }\href {\doibase 10.1088/1475-7516/2016/07/051} {\bibfield
  {journal} {\bibinfo  {journal} {JCAP}\ }\textbf {\bibinfo {volume} {1607}},\
  \bibinfo {pages} {051} (\bibinfo {year} {2016})},\ \Eprint
  {http://arxiv.org/abs/1606.06986} {arXiv:1606.06986 [hep-ph]} \BibitemShut
  {NoStop}%
\bibitem [{\citenamefont {Akita}\ and\ \citenamefont
  {Yamaguchi}(2020)}]{Akita:2020szl}%
  \BibitemOpen
  \bibfield  {author} {\bibinfo {author} {\bibfnamefont {K.}~\bibnamefont
  {Akita}}\ and\ \bibinfo {author} {\bibfnamefont {M.}~\bibnamefont
  {Yamaguchi}},\ }\href@noop {} {\  (\bibinfo {year} {2020})},\ \Eprint
  {http://arxiv.org/abs/2005.07047} {arXiv:2005.07047 [hep-ph]} \BibitemShut
  {NoStop}%
\bibitem [{\citenamefont {Aghanim}\ \emph {et~al.}(2018)\citenamefont {Aghanim}
  \emph {et~al.}}]{Aghanim:2018eyx}%
  \BibitemOpen
  \bibfield  {author} {\bibinfo {author} {\bibfnamefont {N.}~\bibnamefont
  {Aghanim}} \emph {et~al.} (\bibinfo {collaboration} {Planck}),\ }\href@noop
  {} {\  (\bibinfo {year} {2018})},\ \Eprint {http://arxiv.org/abs/1807.06209}
  {arXiv:1807.06209 [astro-ph.CO]} \BibitemShut {NoStop}%
\bibitem [{\citenamefont {Foot}(1991)}]{Foot:1990mn}%
  \BibitemOpen
  \bibfield  {author} {\bibinfo {author} {\bibfnamefont {R.}~\bibnamefont
  {Foot}},\ }\href {\doibase 10.1142/S0217732391000543} {\bibfield  {journal}
  {\bibinfo  {journal} {Mod. Phys. Lett. A}\ }\textbf {\bibinfo {volume} {6}},\
  \bibinfo {pages} {527} (\bibinfo {year} {1991})}\BibitemShut {NoStop}%
\bibitem [{\citenamefont {He}\ \emph {et~al.}(1991)\citenamefont {He},
  \citenamefont {Joshi}, \citenamefont {Lew},\ and\ \citenamefont
  {Volkas}}]{He:1991qd}%
  \BibitemOpen
  \bibfield  {author} {\bibinfo {author} {\bibfnamefont {X.-G.}\ \bibnamefont
  {He}}, \bibinfo {author} {\bibfnamefont {G.~C.}\ \bibnamefont {Joshi}},
  \bibinfo {author} {\bibfnamefont {H.}~\bibnamefont {Lew}}, \ and\ \bibinfo
  {author} {\bibfnamefont {R.}~\bibnamefont {Volkas}},\ }\href {\doibase
  10.1103/PhysRevD.44.2118} {\bibfield  {journal} {\bibinfo  {journal} {Phys.
  Rev. D}\ }\textbf {\bibinfo {volume} {44}},\ \bibinfo {pages} {2118}
  (\bibinfo {year} {1991})}\BibitemShut {NoStop}%
\bibitem [{\citenamefont {Holdom}(1986)}]{Holdom:1985ag}%
  \BibitemOpen
  \bibfield  {author} {\bibinfo {author} {\bibfnamefont {B.}~\bibnamefont
  {Holdom}},\ }\href {\doibase 10.1016/0370-2693(86)91377-8} {\bibfield
  {journal} {\bibinfo  {journal} {Phys. Lett. B}\ }\textbf {\bibinfo {volume}
  {166}},\ \bibinfo {pages} {196} (\bibinfo {year} {1986})}\BibitemShut
  {NoStop}%
\bibitem [{\citenamefont {Ibe}\ \emph {et~al.}(2020)\citenamefont {Ibe},
  \citenamefont {Kobayashi}, \citenamefont {Nakayama},\ and\ \citenamefont
  {Shirai}}]{Ibe:2019gpv}%
  \BibitemOpen
  \bibfield  {author} {\bibinfo {author} {\bibfnamefont {M.}~\bibnamefont
  {Ibe}}, \bibinfo {author} {\bibfnamefont {S.}~\bibnamefont {Kobayashi}},
  \bibinfo {author} {\bibfnamefont {Y.}~\bibnamefont {Nakayama}}, \ and\
  \bibinfo {author} {\bibfnamefont {S.}~\bibnamefont {Shirai}},\ }\href
  {\doibase 10.1007/JHEP04(2020)009} {\bibfield  {journal} {\bibinfo  {journal}
  {JHEP}\ }\textbf {\bibinfo {volume} {04}},\ \bibinfo {pages} {009} (\bibinfo
  {year} {2020})},\ \Eprint {http://arxiv.org/abs/1912.12152} {arXiv:1912.12152
  [hep-ph]} \BibitemShut {NoStop}%
\bibitem [{\citenamefont {Escudero}\ \emph {et~al.}(2019)\citenamefont
  {Escudero}, \citenamefont {Hooper}, \citenamefont {Krnjaic},\ and\
  \citenamefont {Pierre}}]{Escudero:2019gzq}%
  \BibitemOpen
  \bibfield  {author} {\bibinfo {author} {\bibfnamefont {M.}~\bibnamefont
  {Escudero}}, \bibinfo {author} {\bibfnamefont {D.}~\bibnamefont {Hooper}},
  \bibinfo {author} {\bibfnamefont {G.}~\bibnamefont {Krnjaic}}, \ and\
  \bibinfo {author} {\bibfnamefont {M.}~\bibnamefont {Pierre}},\ }\href
  {\doibase 10.1007/JHEP03(2019)071} {\bibfield  {journal} {\bibinfo  {journal}
  {JHEP}\ }\textbf {\bibinfo {volume} {03}},\ \bibinfo {pages} {071} (\bibinfo
  {year} {2019})},\ \Eprint {http://arxiv.org/abs/1901.02010} {arXiv:1901.02010
  [hep-ph]} \BibitemShut {NoStop}%
\bibitem [{\citenamefont {Escudero~Abenza}(2020)}]{Escudero:2020dfa}%
  \BibitemOpen
  \bibfield  {author} {\bibinfo {author} {\bibfnamefont {M.}~\bibnamefont
  {Escudero~Abenza}},\ }\href {\doibase 10.1088/1475-7516/2020/05/048}
  {\bibfield  {journal} {\bibinfo  {journal} {JCAP}\ }\textbf {\bibinfo
  {volume} {05}},\ \bibinfo {pages} {048} (\bibinfo {year} {2020})},\ \Eprint
  {http://arxiv.org/abs/2001.04466} {arXiv:2001.04466 [hep-ph]} \BibitemShut
  {NoStop}%
\bibitem [{\citenamefont {Escudero}(2019)}]{Escudero:2018mvt}%
  \BibitemOpen
  \bibfield  {author} {\bibinfo {author} {\bibfnamefont {M.}~\bibnamefont
  {Escudero}},\ }\href {\doibase 10.1088/1475-7516/2019/02/007} {\bibfield
  {journal} {\bibinfo  {journal} {JCAP}\ }\textbf {\bibinfo {volume} {02}},\
  \bibinfo {pages} {007} (\bibinfo {year} {2019})},\ \Eprint
  {http://arxiv.org/abs/1812.05605} {arXiv:1812.05605 [hep-ph]} \BibitemShut
  {NoStop}%
\bibitem [{\citenamefont {McDermott}\ \emph {et~al.}(2018)\citenamefont
  {McDermott}, \citenamefont {Patel},\ and\ \citenamefont
  {Ramani}}]{McDermott_2018}%
  \BibitemOpen
  \bibfield  {author} {\bibinfo {author} {\bibfnamefont {S.~D.}\ \bibnamefont
  {McDermott}}, \bibinfo {author} {\bibfnamefont {H.~H.}\ \bibnamefont
  {Patel}}, \ and\ \bibinfo {author} {\bibfnamefont {H.}~\bibnamefont
  {Ramani}},\ }\href {\doibase 10.1103/physrevd.97.073005} {\bibfield
  {journal} {\bibinfo  {journal} {Physical Review D}\ }\textbf {\bibinfo
  {volume} {97}} (\bibinfo {year} {2018}),\
  10.1103/physrevd.97.073005}\BibitemShut {NoStop}%
\bibitem [{\citenamefont {Heckler}(1994)}]{Heckler:1994tv}%
  \BibitemOpen
  \bibfield  {author} {\bibinfo {author} {\bibfnamefont {A.~F.}\ \bibnamefont
  {Heckler}},\ }\href {\doibase 10.1103/PhysRevD.49.611} {\bibfield  {journal}
  {\bibinfo  {journal} {Phys. Rev.}\ }\textbf {\bibinfo {volume} {D49}},\
  \bibinfo {pages} {611} (\bibinfo {year} {1994})}\BibitemShut {NoStop}%
\bibitem [{\citenamefont {Fornengo}\ \emph {et~al.}(1997)\citenamefont
  {Fornengo}, \citenamefont {Kim},\ and\ \citenamefont
  {Song}}]{Fornengo:1997wa}%
  \BibitemOpen
  \bibfield  {author} {\bibinfo {author} {\bibfnamefont {N.}~\bibnamefont
  {Fornengo}}, \bibinfo {author} {\bibfnamefont {C.~W.}\ \bibnamefont {Kim}}, \
  and\ \bibinfo {author} {\bibfnamefont {J.}~\bibnamefont {Song}},\ }\href
  {\doibase 10.1103/PhysRevD.56.5123} {\bibfield  {journal} {\bibinfo
  {journal} {Phys. Rev.}\ }\textbf {\bibinfo {volume} {D56}},\ \bibinfo {pages}
  {5123} (\bibinfo {year} {1997})},\ \Eprint
  {http://arxiv.org/abs/hep-ph/9702324} {arXiv:hep-ph/9702324 [hep-ph]}
  \BibitemShut {NoStop}%
\bibitem [{\citenamefont {Mangano}\ \emph {et~al.}(2002)\citenamefont
  {Mangano}, \citenamefont {Miele}, \citenamefont {Pastor},\ and\ \citenamefont
  {Peloso}}]{Mangano:2001iu}%
  \BibitemOpen
  \bibfield  {author} {\bibinfo {author} {\bibfnamefont {G.}~\bibnamefont
  {Mangano}}, \bibinfo {author} {\bibfnamefont {G.}~\bibnamefont {Miele}},
  \bibinfo {author} {\bibfnamefont {S.}~\bibnamefont {Pastor}}, \ and\ \bibinfo
  {author} {\bibfnamefont {M.}~\bibnamefont {Peloso}},\ }\href {\doibase
  10.1016/S0370-2693(02)01622-2} {\bibfield  {journal} {\bibinfo  {journal}
  {Phys. Lett.}\ }\textbf {\bibinfo {volume} {B534}},\ \bibinfo {pages} {8}
  (\bibinfo {year} {2002})},\ \Eprint {http://arxiv.org/abs/astro-ph/0111408}
  {arXiv:astro-ph/0111408 [astro-ph]} \BibitemShut {NoStop}%
\bibitem [{\citenamefont {Chacko}\ \emph {et~al.}(2004)\citenamefont {Chacko},
  \citenamefont {Hall}, \citenamefont {Okui},\ and\ \citenamefont
  {Oliver}}]{Chacko:2003dt}%
  \BibitemOpen
  \bibfield  {author} {\bibinfo {author} {\bibfnamefont {Z.}~\bibnamefont
  {Chacko}}, \bibinfo {author} {\bibfnamefont {L.~J.}\ \bibnamefont {Hall}},
  \bibinfo {author} {\bibfnamefont {T.}~\bibnamefont {Okui}}, \ and\ \bibinfo
  {author} {\bibfnamefont {S.~J.}\ \bibnamefont {Oliver}},\ }\href {\doibase
  10.1103/PhysRevD.70.085008} {\bibfield  {journal} {\bibinfo  {journal} {Phys.
  Rev. D}\ }\textbf {\bibinfo {volume} {70}},\ \bibinfo {pages} {085008}
  (\bibinfo {year} {2004})},\ \Eprint {http://arxiv.org/abs/hep-ph/0312267}
  {arXiv:hep-ph/0312267} \BibitemShut {NoStop}%
\bibitem [{\citenamefont {Escudero}\ and\ \citenamefont
  {Witte}(2020)}]{Escudero:2019gvw}%
  \BibitemOpen
  \bibfield  {author} {\bibinfo {author} {\bibfnamefont {M.}~\bibnamefont
  {Escudero}}\ and\ \bibinfo {author} {\bibfnamefont {S.~J.}\ \bibnamefont
  {Witte}},\ }\href {\doibase 10.1140/epjc/s10052-020-7854-5} {\bibfield
  {journal} {\bibinfo  {journal} {Eur. Phys. J. C}\ }\textbf {\bibinfo {volume}
  {80}},\ \bibinfo {pages} {294} (\bibinfo {year} {2020})},\ \Eprint
  {http://arxiv.org/abs/1909.04044} {arXiv:1909.04044 [astro-ph.CO]}
  \BibitemShut {NoStop}%
\bibitem [{\citenamefont {Semikoz}\ and\ \citenamefont
  {Tkachev}(1995)}]{Semikoz:1994zp}%
  \BibitemOpen
  \bibfield  {author} {\bibinfo {author} {\bibfnamefont {D.}~\bibnamefont
  {Semikoz}}\ and\ \bibinfo {author} {\bibfnamefont {I.}~\bibnamefont
  {Tkachev}},\ }\href {\doibase 10.1103/PhysRevLett.74.3093} {\bibfield
  {journal} {\bibinfo  {journal} {Phys. Rev. Lett.}\ }\textbf {\bibinfo
  {volume} {74}},\ \bibinfo {pages} {3093} (\bibinfo {year} {1995})},\ \Eprint
  {http://arxiv.org/abs/hep-ph/9409202} {arXiv:hep-ph/9409202} \BibitemShut
  {NoStop}%
\bibitem [{\citenamefont {Semikoz}\ and\ \citenamefont
  {Tkachev}(1997)}]{Semikoz:1995rd}%
  \BibitemOpen
  \bibfield  {author} {\bibinfo {author} {\bibfnamefont {D.}~\bibnamefont
  {Semikoz}}\ and\ \bibinfo {author} {\bibfnamefont {I.}~\bibnamefont
  {Tkachev}},\ }\href {\doibase 10.1103/PhysRevD.55.489} {\bibfield  {journal}
  {\bibinfo  {journal} {Phys. Rev. D}\ }\textbf {\bibinfo {volume} {55}},\
  \bibinfo {pages} {489} (\bibinfo {year} {1997})},\ \Eprint
  {http://arxiv.org/abs/hep-ph/9507306} {arXiv:hep-ph/9507306} \BibitemShut
  {NoStop}%
\bibitem [{\citenamefont {Pitrou}\ \emph {et~al.}(2018)\citenamefont {Pitrou},
  \citenamefont {Coc}, \citenamefont {Uzan},\ and\ \citenamefont
  {Vangioni}}]{Pitrou:2018cgg}%
  \BibitemOpen
  \bibfield  {author} {\bibinfo {author} {\bibfnamefont {C.}~\bibnamefont
  {Pitrou}}, \bibinfo {author} {\bibfnamefont {A.}~\bibnamefont {Coc}},
  \bibinfo {author} {\bibfnamefont {J.-P.}\ \bibnamefont {Uzan}}, \ and\
  \bibinfo {author} {\bibfnamefont {E.}~\bibnamefont {Vangioni}},\ }\href
  {\doibase 10.1016/j.physrep.2018.04.005} {\bibfield  {journal} {\bibinfo
  {journal} {Phys. Rept.}\ }\textbf {\bibinfo {volume} {754}},\ \bibinfo
  {pages} {1} (\bibinfo {year} {2018})},\ \Eprint
  {http://arxiv.org/abs/1801.08023} {arXiv:1801.08023 [astro-ph.CO]}
  \BibitemShut {NoStop}%
\bibitem [{\citenamefont {Fields}\ \emph {et~al.}(2020)\citenamefont {Fields},
  \citenamefont {Olive}, \citenamefont {Yeh},\ and\ \citenamefont
  {Young}}]{Fields:2019pfx}%
  \BibitemOpen
  \bibfield  {author} {\bibinfo {author} {\bibfnamefont {B.~D.}\ \bibnamefont
  {Fields}}, \bibinfo {author} {\bibfnamefont {K.~A.}\ \bibnamefont {Olive}},
  \bibinfo {author} {\bibfnamefont {T.-H.}\ \bibnamefont {Yeh}}, \ and\
  \bibinfo {author} {\bibfnamefont {C.}~\bibnamefont {Young}},\ }\href
  {\doibase 10.1088/1475-7516/2020/03/010} {\bibfield  {journal} {\bibinfo
  {journal} {JCAP}\ }\textbf {\bibinfo {volume} {03}},\ \bibinfo {pages} {010}
  (\bibinfo {year} {2020})},\ \Eprint {http://arxiv.org/abs/1912.01132}
  {arXiv:1912.01132 [astro-ph.CO]} \BibitemShut {NoStop}%
\bibitem [{\citenamefont {Berlin}\ \emph {et~al.}(2019)\citenamefont {Berlin},
  \citenamefont {Blinov},\ and\ \citenamefont {Li}}]{Berlin:2019pbq}%
  \BibitemOpen
  \bibfield  {author} {\bibinfo {author} {\bibfnamefont {A.}~\bibnamefont
  {Berlin}}, \bibinfo {author} {\bibfnamefont {N.}~\bibnamefont {Blinov}}, \
  and\ \bibinfo {author} {\bibfnamefont {S.~W.}\ \bibnamefont {Li}},\ }\href
  {\doibase 10.1103/PhysRevD.100.015038} {\bibfield  {journal} {\bibinfo
  {journal} {Phys. Rev. D}\ }\textbf {\bibinfo {volume} {100}},\ \bibinfo
  {pages} {015038} (\bibinfo {year} {2019})},\ \Eprint
  {http://arxiv.org/abs/1904.04256} {arXiv:1904.04256 [hep-ph]} \BibitemShut
  {NoStop}%
\end{thebibliography}%

\end{document}